\documentclass[12pt]{article}

\usepackage[ascii]{inputenc}
\usepackage{amsmath,amssymb,amsfonts,amsthm}
\usepackage[margin=2.8cm]{geometry}
\usepackage{graphicx}
\usepackage[caption=false]{subfig}
\usepackage[pdftex,bookmarks=false,colorlinks=true,linkcolor=blue,
citecolor=blue,filecolor=black,urlcolor=blue]{hyperref}

\usepackage{latexsym}
\usepackage{longtable}
\usepackage{epsfig}
\usepackage{psfrag}
\numberwithin{equation}{section}

\begin{document}
\begin{center}
\vspace*{2cm} {\Large {\bf Fluctuating hydrodynamics approach to equilibrium time correlations for anharmonic chains  \bigskip\bigskip\\}}
{\large Herbert Spohn}\bigskip\bigskip\\
Zentrum Mathematik and Physik Department,
Technische Universit\"at M\"unchen,
Boltzmannstra{\ss}e 3, 85747 Garching, Germany. 
{\tt spohn@ma.tum.de}
\vspace{5cm}\end{center}
\textbf{Abstract.} Linear fluctuating hydrodynamics is a useful and versatile  tool for describing fluids, as well as other systems with conserved fields, on a mesoscopic scale. In one spatial dimension, however, transport is anomalous, which requires to develop 
a nonlinear extension of fluctuating hydrodynamics. The relevant nonlinearity turns out to be the quadratic part of the Euler currents when expanding relative to a uniform background. We outline the theory and compare with recent
molecular dynamics simulations. 
\newpage\noindent
\textbf{Table of contents}
\begin{enumerate}
\item Introduction, long time tails for simple fluids
\item Anharmonic chains\\
2.1 Conservation laws and equilibrium time correlations\\
2.2 Linearized hydrodynamics
\item Nonlinear fluctuating hydrodynamics\\
3.1 Euler currents to second order\\
3.2 Stationary measure for the physical fields\\
3.3 Transformation to normal modes
\item Mode-coupling theory\\
4.1 Decoupling hypothesis\\
4.2 One-loop, diagonal, and small overlap approximations\\
4.3 Numerical simulations of the mode-coupling equations\\
4.4 Asymptotic self-similarity\\
4.5 No signal beyond the sound cone\\
4.6 Dynamical phase diagram
\item Molecular dynamics simulations
\item Total current correlations
\item  Other Hamiltonian chains\\
7.1 Coupled rotators\\
7.2 Lattice nonlinear Schr\"{o}dinger equation
\end{enumerate}\newpage\noindent

\section{Introduction, long time tails for simple fluids}
\label{sec1}
 
In the mid 1950ies, Green \cite{Gr54} and Kubo \cite{Ku57,Ku59} discovered that transport coefficients for simple fluids can be obtained through a time-integral over the respective total current correlation function. For tracer diffusion such a connection is more immediate and was understood much earlier. But the then novel insight was that collective transport coefficients, such as viscosity and thermal conductivity, follow the same pattern. Thus it became a central issue to determine the time decay of such current correlations. With essentially no tools available this amounted to an impossible task.  The static equilibrium correlations were known to decay exponentially fast,  as confirmed by a convergent series expansion. But for the dynamics one would have to deal with a huge set of coupled differential equations. At the time the only theoretical tool available
 was the Boltzmann equation valid at low density. Kinetic theory predicts an exponential decay for the current time correlations and it was tacitly assumed that such behavior would extend to moderate densities. Alder and Wainwright \cite{AlWa70}
 tried to check the situation in a
 pioneering molecular dynamics simulation of  500 hard disks, resp. hard spheres, at periodized volume 2, 3, and 5 times larger than close packing. For tracer diffusion  they convincingly observed a power law decay as $t^{-d/2}$, dimension $d= 2,3$, which was baptized  ``long time tail''. They also argued that the same behavior should hold for collective transport. Theory quickly jumped in and predicted 
 a decay as $t^{-d/2}$ for viscosity and thermal conductivity \cite{PoRe75,ErHa76}.
 
 There are several theoretical schemes and they all arrive at the same prediction, which of course increases their confidence level. As recognized for some time, the most direct approach is linear fluctuating hydrodynamics plus small nonlinear perturbations. We refer to the recent monograph \cite{OrSe06} for a comprehensive discussion. Here I provide only a rough sketch of the method with the purpose  to explain why one dimension is so special. In the physical dimension  a simple fluid has five conservation laws and correspondingly fluctuating hydrodynamics has to deal with a five component field, where the momentum components are odd, density and energy are even under time reversal. As well known
 \cite{ReDe77,OrSe06}, the full structure has to be used  in order to arrive at quantitative predictions. But the argument becomes even more direct for the 
 (unphysical) case of a single conservation law.
 
 Let us thus consider the scalar field $\rho(\boldsymbol{x},t)$, which for concreteness is called density.  Space is $\boldsymbol{x} \in \mathbb{R}^d$
 and time is $t \in \mathbb{R}$. $\rho(\boldsymbol{x},t)$ is a fluctuating
 field. On the macroscopic scale fluctuations are not visible and $\rho$ satisfies the hyperbolic conservation law
\begin{equation}\label{1.1}
\partial_t \rho +\nabla \cdot\vec{j}(\rho) = 0\,,
\end{equation}
where $\vec{j}$ is the density current. To also include mesoscopic details, in particular to incorporate fluctuations, one argues that the current
has, in addition to the deterministic part, also a random contribution which is essentially uncorrelated in space-time. Since fluctuations are always associated with dissipation,  
on a more refined scale Eq. \eqref{1.1} becomes
\begin{equation}\label{1.2}
\partial_t \rho +\nabla\cdot \big(\vec{j}(\rho) -D \nabla \rho + \sigma \vec{\xi}\,\big) = 0\,.
\end{equation}
Here $\vec{\xi}(\boldsymbol{x},t)$ is Gaussian white noise with mean zero and covariance
\begin{equation}\label{1.3}
\langle \xi_\alpha(\boldsymbol{x},t)\xi_{\alpha'}(\boldsymbol{x}',t')\rangle =  \delta_{\alpha\alpha'}\delta(\boldsymbol{x}-\boldsymbol{x}')\delta(t - t')\,,\quad \alpha,\alpha' = 1,...,d\,.
\end{equation}
$\sigma$ is the noise strength and $D$ is the diffusion constant. They are both treated as numbers.
Physically they will depend on the density. But this would be higher order effects, which are ignored in our discussion. 

The goal is to compute the density time correlations in  the stationary regime, which is no easy task, since \eqref{1.2} is nonlinear. But correlations can be thought of as imposing at $t = 0$ a small density perturbation  at the origin  and then 
record how the perturbation propagates in space-time. For this purpose one might hope to get away with linearizing \eqref{1.2}
at the uniform background density $\rho_0$ as $\rho_ 0+ \varrho(\boldsymbol{\boldsymbol{x}},t)$, which yields
\begin{equation}\label{1.4}
\partial_t \varrho +\nabla \cdot\big(\vec{j}'(\rho_0)\varrho - D \nabla \varrho + \sigma \vec{\xi}\,\big) = 0\,,
\end{equation}
where $'$ refers to differentiation w.r.t $\rho$. \eqref{1.4} is a linear Langevin equation, hence solved easily. Since $\varrho$ is a deviation from the uniform background, we are interested in the space-time stationary process with zero mean. First  note that \eqref{1.4} has a unique time-stationary zero mean measure, which  
is Gaussian white noise in the spatial variable,
\begin{equation}\label{1.5}
\langle \varrho(\boldsymbol{x}) \varrho(\boldsymbol{x}')\rangle = (\sigma^2/2D)\delta(\boldsymbol{x} - \boldsymbol{x}')\,.
\end{equation}
For the stationary space-time covariance one obtains 
\begin{equation}\label{1.6}
\langle \varrho(\boldsymbol{x},t) \varrho(0,0)\rangle = (\sigma^2/2D)p(\boldsymbol{x} - \vec{c}\,t,t)\,,
\end{equation} 
where $p$ is the Gaussian transition kernel,
\begin{equation}\label{1.6a}
p(\boldsymbol{x},t) = (4\pi D |t|)^{-d/2}\exp\big(- \boldsymbol{x}^2/4 D |t|\big) \,,
\end{equation} 
 and $\vec{c} = \vec{j}'(\rho_0)$. By \eqref{1.4} the fluctuating current is given by
\begin{equation}\label{1.7}
\vec{\mathcal{J}} = \vec{c}\varrho - D \nabla \varrho + \sigma \vec{\xi}\,.
\end{equation}
For the stationary correlations of the total current one arrives at
\begin{equation}\label{1.8}
\int_{\mathbb{R}^d} dx \langle \vec{\mathcal{J}}(\boldsymbol{x},t) \cdot\vec{\mathcal{J}}(0,0)\rangle = d\sigma^2 \delta(t)\,.
\end{equation}

No surprise. A density fluctuation propagates with velocity $\vec{c}$ and spreads diffusively. The currents are delta-correlated,
which should translate into exponential decay for the underlying microscopic system. But before jumping at such conclusions one has to study the stability of  \eqref{1.8} against including higher orders in the expansion. By power counting the next to leading term is the nonlinear current at second order, which amounts to
\begin{equation}\label{1.9}
\partial_t \varrho +\nabla \cdot\big(\vec{c}\varrho + \vec{G} \varrho^2-D \nabla \varrho + \sigma \vec{\xi}\,\big) = 0\,, \quad  \vec{G} = \tfrac{1}{2} \vec{j}\,''(\rho_0)\,.
\end{equation}
The task is to compute the current correlation  \eqref{1.8} perturbatively in $\vec{G}$.

We first remove $\vec{c}$ by switching to a moving frame of reference. Secondly we note that, quite surprisingly, white noise is still time-stationary
under the nonlinear Langevin equation \eqref{1.9}. Since, as argued above,  spatial white noise is  time-stationary under the linear part, one only has to check its
invariance under the deterministic evolution $\partial_t \varrho =- \vec{G}\cdot \nabla\varrho^2$. Formally its vector field  is divergence free, since one can choose a divergence free discretization \cite{SaSp09}, compare with the discussion in Section \ref{sec3} below. Hence the ``volume measure" is preserved and one has to check only the time change of the logarithm of the stationary density,
\begin{equation}\label{1.10}
\frac{d}{dt} \int_{\mathbb{R}^d} dx  \tfrac{1}{2}\varrho(\boldsymbol{x})^2 = -\int_{\mathbb{R}^d} dx\varrho(\boldsymbol{x})\vec{G}\cdot \nabla\varrho^2(\boldsymbol{x}) =  \tfrac{1}{3}\int_{\mathbb{R}^d} dx\vec{G}\cdot \nabla\varrho^3(\boldsymbol{x})= 0 \,.
\end{equation}

For an expansion in $\vec{G}\cdot \nabla\varrho^2$ it is most convenient to use the Fokker-Planck generator, denoted by $L = L_0 +L_1$, where $L_0$
corresponds to the Gaussian part and $L_1$ to the nonlinear flow. We define
\begin{equation}\label{1.11}
S(\boldsymbol{x},t) = \langle \varrho(\boldsymbol{x},t) \varrho(0,0)\rangle  = \langle \varrho(\boldsymbol{x}) \mathrm{e}^{Lt}\varrho(0)\rangle\,,
\end{equation}
average with respect to the time-stationary Gaussian measure, and plan to use the general second moment sum rule
\begin{equation}\label{1.16}
\frac{d^2}{dt^2}\int_{\mathbb{R}^d} dx  \boldsymbol{x}^2 S(\boldsymbol{x},t) =  \int_{\mathbb{R}^d} dx \langle  \vec{\mathcal{J}}(\boldsymbol{x},t) \cdot \vec{\mathcal{J}}(0,0) \rangle\,, 
\end{equation}
which follows directly from the conservation law.
By second order time-dependent perturbation theory,
\begin{eqnarray}\label{1.12a}
&&\hspace{-26pt}S(\boldsymbol{x},t) = \langle \varrho(\boldsymbol{x}) e^{L_{0}t}\varrho(0)\rangle +  \int _0^t dt_1\langle \varrho(\boldsymbol{x}) \mathrm{e}^{L_{0}(t-t_1)}L_1\mathrm{e}^{L_{0}t_1}\varrho(0)\rangle   \nonumber\\
&&\hspace{20pt} + \int _0^t dt_2 \int _0^{t_2}dt_1 \langle \varrho(\boldsymbol{x}) \mathrm{e}^{L_{0}(t-t_2)}L_1\mathrm{e}^{L_{0}(t_2-t_1)}L_1
\mathrm{e}^{L_0t_1}\varrho(0)\rangle + \mathcal{O}(|\vec{G}|^3)\,.
\end{eqnarray}
 $L_0\varrho$ is linear, while 
$L_1\varrho$ is quadratic in $\varrho$. Thus the second term on the right is odd in $\varrho$ and vanishes. 
For any functional, $F$,
 \begin{equation}\label{1.13}
\langle \varrho(\boldsymbol{x})  L_1 F(\varrho)\rangle = - \langle (L_1 \varrho (\boldsymbol{x}))  F(\varrho)\rangle\,,
\end{equation}
see \eqref{B.18a}, and the left $L_1$ is swapped over to act on $\mathrm{e}^{L_{0}(t-t_2)}  \varrho(\boldsymbol{x})$. By translation invariance,
 $\vec{G}\cdot\nabla$ can be pulled in front as acting on $\boldsymbol{x}$. Combining the terms one arrives at
 \begin{eqnarray}\label{1.14}
&&\hspace{-28pt} S(\boldsymbol{x},t) =   \langle \varrho(\boldsymbol{x}) \mathrm{e}^{L_{0}t} \varrho(0)\rangle 
 +   
  \int _0^t dt_2 \int _0^{t_2}dt_1(\vec{G}\cdot\nabla)^2\nonumber\\
 &&\hspace{20pt} \times \int_{\mathbb{R}^d} dx_1\int_{\mathbb{R}^d} dx_2
 p(\boldsymbol {x} - \boldsymbol {x}_2,t - t_2) p(\boldsymbol {x}_1,t_1)
  \langle\varrho(\boldsymbol{x}_2)^2  \mathrm{e}^{L_{0}(t_2 -t_1)}\varrho(\boldsymbol {x}_1)^2\rangle \,.
 \end{eqnarray}
The Gaussian average is computed as
\begin{equation}\label{1.14a}
 \langle\varrho(\boldsymbol{x}_2)^2  \mathrm{e}^{L_{0}t}\varrho(\boldsymbol {x}_1)^2\rangle
 =   2 \big((\sigma^2/2D)p(\boldsymbol{x}_2 - \boldsymbol{x}_1,t)\big)^2 +\, \mathrm{s\mbox{-}c}\,.
\end{equation}
 The self-contraction does not depend on $\boldsymbol{x}$, hence vanishes when applying $\vec{G}\cdot\nabla$.  We insert in \eqref{1.16}.  Working out the integrals yields, including second order, 
 \begin{equation}\label{1.17}
\int_{\mathbb{R}^d} dx \langle   \vec{\mathcal{J}}(\boldsymbol{x},t) \cdot \vec{\mathcal{J}}(0,0) \rangle = d\sigma^2\delta(t) + 4(\sigma^2/2D)^2|\vec{G}|^2 p(0,2t)\,,
\end{equation}
which is the claimed long time tail for a scalar conserved field in $d$ dimensions. 

Eq. \eqref{1.9} is singular at short distances, the worse the higher the dimension. In physical systems there is a natural cut-off at the microscopic scale which is simply ignored in \eqref{1.9}. There are many possibilities to improve, but the serious constraint
is to obtain a still manageable nonlinear stochastic equation. The noise should remain $\delta$-correlated in time so to preserve the Markov property of the time  evolution. One could smoothen in space, but thereby may loose the information on the time-stationary measure.
To my experience the best compromise is to adopt the obvious spatial discretization. Then one has a set of coupled stochastic differential equations. For them one can check rigorously the time-stationary measure and  identities as 
\textit{e.g.}  \eqref{1.13}. 
On the perturbative level this then leads to the continuum expressions as \eqref{1.14}. For a simple fluid the current correlations are bounded and will not diverge near $t=0$. So in \eqref{1.17} only the long time prediction can be trusted.

What can be learned from the long time tails? In dimension $d \geq 3$ the decay is integrable. Thus, as assumed implicitly
beforehand, the model has a finite diffusivity. Higher order terms in the expansion will modify prefactors but should not alter the exponent for the time decay. Dimension $d=2$ is marginal. The diffusivity is weakly divergent. In principle, say, a system of hard disks 
has infinite viscosity and thermal conductivity. But since the divergence is only logarithmic one could convert the result into a weak system size dependence. In \textit{one dimension} the conductivity is truly infinite. Obviously, not even the power law as based on the perturbative expansion \eqref{1.14} can be trusted and one needs to develop non-perturbative techniques.

Eq. \eqref{1.9}  for $d=1$ is known as stochastic Burgers equation, which we record for later reference as
\begin{equation}\label{1.18a}
\partial_t \varrho -\partial_x\big( \tfrac{1}{2}\lambda \varrho^2  +\nu \partial_x \varrho +\sigma \xi\big) = 0\,,\quad x \in \mathbb{R}\,.
\end{equation}
One can introduce a potential through
$\varrho = \partial_x h$. Then $h$ satisfies the one-dimensional version of the Kadar-Parisi-Zhang (KPZ)
equation \cite{KPZ86},
\begin{equation}\label{1.18}
\partial_t h = \tfrac{1}{2}\lambda (\partial_x h)^2  +\nu \partial_x^2 h +\sigma \xi\,,
\end{equation}
for the moment using the more conventional symbols for the coefficients. 
Over the last fifteen years many properties, including rigorous results, of the KPZ equation have been obtained. While this is not the place to dive into details,  I note that the solution is continuous in $x$, but so singular that 
$(\partial_x h)^2$ is ill-defined. However it is proved that the ultraviolet divergence is very mild and can be tamed by what would be an infinite energy renormalization in a 1+1 dimensional quantum field theory, compare with the discussion in \cite{Ku14}. More precisely one chooses a regularized version of \eqref{1.18} by
  replacing $\xi(x,t)$ through the spatially smoothed version
$\xi_\varphi(x,t) = \int \varphi(x - x')\xi(x',t)dx'$ with $\varphi \geq 0$, even, smooth, of rapid decay at infinity, and normalized to $1$. It can be proved  that then the solution to \eqref{1.18}
is well-defined. One introduces an ultraviolet cut-off of spatial size $\epsilon$ by choosing the $\delta$-function sequence  $\varphi_\epsilon(x) = \epsilon^{-1}\varphi(\epsilon^{-1}x)$ and  substituting the noise $\xi$ by its smoothed version $\xi_\epsilon = \xi_{\varphi_\epsilon}$. Let us denote the corresponding solution of the KPZ equation by $h_\epsilon(x,t)$. Then $h_\epsilon(x,t) - v_\epsilon t$, \textit{i.e.} $h_\epsilon(x,t)$ viewed in the  frame moving with the diverging velocity $v_\epsilon  = \epsilon^{-1}\int \varphi(x)^2dx$, has a non-degenerate limit as $\epsilon \to 0$, which coincides with the Cole-Hopf solution of the KPZ equation \cite{BC,BeGi97}.

In this context Hairer recently developed a solution theory, for which he was awarded the 2014 Fields Medal.
His theory works for the KPZ equation, as well as a large class of other singular stochastic partial differential equations,   and for general cutoffs \cite{Ha13,Ha13a}.  Of course,  the solution theory studies the small scale structure of solutions, and not the large scale, where the universal behaviors of interest are observed.

There is an exact  formula for $S(x,t)$, which involves Fredholm determinants. In the long time limit
\begin{equation}\label{1.19}
S(x,t) \simeq (\sigma^2/2\nu) (\sqrt{2}\lambda |t|)^{-2/3}f_\mathrm{KPZ}\big(\sqrt{2}\lambda |t|)^{-2/3}x\big)\,.
\end{equation}
The scaling function $f_\mathrm{KPZ}$ will reappear below, where more details are given. The Burgers current  reads 
\begin{equation}\label{1.19a}
\mathcal{J}(x,t) =  -( \tfrac{1}{2}\lambda \varrho^2  +\nu \partial_x \varrho +\sigma \xi\big)
\end{equation}
and for its total  correlation function one obtains, using the sum rule \eqref{1.16},
\begin{equation}\label{1.20}
\int_\mathbb{R} dx \langle   \mathcal{J}(x,t)  \mathcal{J}(0,0) \rangle \simeq \Big((\sigma^2/2\nu) (\sqrt{2}\lambda)^2
\int_\mathbb{R} dx \tfrac{4}{9}x^2 f_\mathrm{KPZ}(x)\Big) (\sqrt{2}\lambda t)^{-2/3}
\end{equation}
valid for large $t$. Note that the true decay turns out to be $ t^{-2/3} $, to be contrasted with the perturbative result $t^{-1/2}$. In fact the power $2/3$ was argued already in \cite{PoRe75}  using a self-consistent scheme, see also
\cite{FoNe77}. As a fairly unusual feature,  the non-universal coefficients are given directly in terms of the parameters of the stochastic Burgers equation. In fact, the particular form can be guessed from the scaling properties of Eq. \eqref{1.18a}. Only for the scaling function $f_\mathrm{KPZ}$ and the pure number $\sqrt{2}$ one has to rely on the exact solution, which is the  result of an intricate analysis using lattice type models \cite{PrSp04,FeSp06,BaFePe10,FeSpWe15}, replica computations for the KPZ equation \cite{ImSa12,ImSa13}, and the finite time exact solution of the stationary KPZ equation itself \cite{BoCoFeV14}.

In these notes we will explain how nonlinear fluctuating hydrodynamics, in the same spirit as already explained for a scalar field in one dimension, can be used to predict the asymptotic form of the equilibrium space-time correlations of anharmonic chains. Chains are one-dimensional objects and it may seem as if we have accomplished the task already. Well, we have not even written down a Hamiltonian. So first we have to dwell 
on a few general properties of anharmonic chains. In particular we will see that they have three conserved fields, generically. 
The corresponding Euler equations are derived, thereby identifying the macroscopic currents. But now we are forced to handle 
several conserved fields. While it is not so difficult to write down the multi-component generalization of \eqref{1.18a},
the analysis of the Langevin equation will be more complicated with no exact solution at help.
A second major task will be to test the quality of these predictions by comparing with molecular dynamics simulations.\\\\
\textbf{Acknowledgements}. First of all I owe my thanks to Christian Mendl. Only through his constant interest, his help in checking details, and his superb numerical efforts my project could advance to its current level.  Various collaborations developed and I am most grateful to my coauthors A. Dhar, P. Ferrari,  D. Huse, M. Kulkarni, T. Sasamoto, and G. Stoltz. At various stages of the project, I greatly benefited from discussions with
H. van Beijeren, C. Bernardin, J. Krug, J.L. Lebowitz, S. Lepri, R. Livi, S. Olla, H. Posch, and G. Sch\"{u}tz. 
 \section{Anharmonic chains}\label{sec2}
 \textbf{Conservation laws and equilibrium time correlations}. We consider a classical fluid consisting of particles with positions $q_j$ and momenta $p_j$, $j = 1,...,N$, $q_j,p_j \in \mathbb{R}$,
possible boundary conditions to be discussed later on. We use units such that the mass of the particles equals 1. Then the Hamiltonian is of the standard form,
\begin{equation}\label{2.1}
H_N^\mathrm{fl} = \sum_{j=1}^N \tfrac{1}{2} p_j^2 + \tfrac{1}{2}\sum_{i \neq j = 1}^NV(q_i - q_j)\,, 
\end{equation}
with pair potential $V(x) = V(-x)$. The potential may have a hard core and otherwise is assumed to be short ranged.
The dynamics for long range potentials is of independent interest \cite{Ru14}, but not discussed here. Furthermore the potential is assumed to be thermodynamically stable, meaning the validity of a bound as $\sum_{i \neq j = 1}^NV(q_i - q_j) \geq A - BN$ for some constants
$A$ and $B >0$.  A substantial simplification is achieved by assuming a hard core of diameter $a$, \textit{i.e.}  $V(x) = \infty$ for $|x| < a$, and restricting the range of the smooth part of the potential to at most $2a$. Then the particles maintain their order,  $q_j \leq q_{j+1}$, and in addition only nearest neighbor particles interact. Hence $H_N^\mathrm{fl}$
simplifies to
\begin{equation}\label{2.2}
H_N = \sum_{j=1}^N  \tfrac{1}{2}p_j^2 + \sum_{j = 1}^{N-1}V(q_{j+1} - q_j)\,.
\end{equation}
As a, at first sight very different,  physical realization,  we could interpret $H_N$ as describing particles in one dimension coupled through anharmonic springs which is then usually referred to as anharmonic chain.

In the second interpretation the spring potential can be more general than anticipated so far. No ordering constraint is  required and the potential does not have to be even. To have
well defined thermodynamics the chain is pinned at both ends as $q_1 =0$ and $q_{N+1} = \ell N$. It is convenient to introduce the stretch $r_j = q_{j+1} - q_j$. Then the boundary condition corresponds to the microcanonical constraint
\begin{equation}\label{2.3}
\sum_{j=1}^{N} r_j = \ell N\,. 
\end{equation} 
Switching   to canonical equilibrium according to the standard rules, one the arrives at the  obvious condition  of 
a finite partition function
\begin{equation}\label{2.3a}
Z(P,\beta) =  \int dx \,\mathrm{e}^{-\beta(V(x) +Px)} < \infty \,, 
\end{equation} 
using the standard convention that the integral is over the entire real line. Here $\beta  > 0 $ is the inverse temperature and $P$ is the thermodynamically conjugate variable to the stretch.
By partial integration
\begin{equation}\label{2.4}
 P = -Z(P,\beta)^{-1} \int dx V'(x)\,\mathrm{e}^{-\beta(V(x) +Px)}  \,, 
\end{equation} 
implying that $P$ is the average force in the spring between two adjacent particles, hence identified as thermodynamic  pressure. To have a finite partition function, a natural condition on the potential is to be bounded from below 
and to have a one-sided linear bound as 
$V(x) \geq a_0 + b_0|x|$ for either $x> 0$ or $x < 0$ and $b_0 >0$. Then there is a non-empty interval $I(\beta)$ such that $Z(P,\beta) < \infty $ for $P\in I(\beta)$. For the particular case of  a hard-core fluid one imposes  $P >0$.
\smallskip\\
\textit{Note:} The sign of $P$ is chosen such that for a gas of hard-point particles one has the familiar ideal gas law $P = 1/\beta \ell$. The chain tension is $-P$. \smallskip

Famous examples are the harmonic chain, $V_\mathrm{ha}(x) = x^2$, the Fermi-Pasta-Ulam (FPU) chain,
$V_\mathrm{FPU}(x) = \tfrac{1}{2}x^2 + \tfrac{1}{3}\alpha x^3 + \tfrac{1}{4}\beta x^4$, in the historical notation
\cite{FPU56}, and the Toda chain \cite{To60}, $V(x) = \mathrm{e}^{-x}$, in which case $P>0$ is required. The harmonic chain, the Toda chain, and the hard-core potential, $V_\mathrm{hc}(x) = \infty$ for $|x| < a$ and $V_\mathrm{hc}(x) = 0$ for $|x| \geq a$, are in fact integrable systems which have a very different correlation structure and will not be discussed here. Except for the harmonic chain, one simple way to break integrability is to assume alternating masses, say $m_j = m_0$ for even $j$ and $m_j = m_1$ for odd $j$.

We will mostly deal with anharmonic chains described by the Hamiltonian \eqref{2.2}, including  one-dimensional hard-core fluids with a sufficiently small potential range. There are several good reasons. Firstly on the level of fluctuating hydrodynamics a generic one-dimensional fluid   cannot be distinguished from an anharmonic chain. Thus with the proper translation of the various terms we would also predict the large scale correlation structure of one-dimensional fluids.  
The second reason is that in the large body of molecular dynamics simulations there is not a single one which deals
with an ``honest'' one-dimensional fluid. To be able to reach large system sizes all simulations are performed for anharmonic chains. 
In addition, from a theoretical perspective, the equilibrium measures of anharmonic chains are particularly simple in being
of product form in momentum and stretch variables. Thus material parameters, as compressibility and sound speed, can be
expressed in terms of one-dimensional integrals involving the Boltzmann factor $\mathrm{e}^{-\beta(V(x) +Px)}$, $V(x)$,
and $x$.

Anharmonic chains should be thought of as a particular class of 1+1 dimensional field theories. Thus $q_j$ is viewed as the displacement variable
at lattice site $j$ and not necessarily the physical position of the $j$-th particle on the real line. There is a simple translation
between both pictures, but we will stick to the field theory point of view. For a fluid with unlabeled particles and, say, a bounded potential, the equivalence is lost and only the fluid picture can be used.

This being said, we follow the standard rules. We write down the dynamics of stretches and momenta and  identify the conserved  fields. From there we infer the microcanonical and canonical equilibrium measures. For slowly varying equilibrium parameters we deduce the Euler equations. In particular, their version linearized at uniform equilibrium will
constitute the backbone in understanding the equilibrium time correlations of the conserved fields.

The dynamics of the anharmonic chain is governed by 
\begin{equation}\label{2.6}
\frac{d}{dt} q_j=p_{j}\,,\qquad \frac{d}{dt}{p}_j=V'(q_{j+1} - q_j) - V'(q_j - q_{j-1})\,.
\end{equation}
For the initial conditions we choose a lattice cell of length $N$ and require
\begin{equation}\label{2.6a}
q_{j+N} = q_j +\ell N\,,\qquad p_{j+N} = p_j
\end{equation}
for all $j \in \mathbb{Z}$. This property is preserved under the dynamics and thus properly mimics a system of finite length $N$. The stretches are then $N$-periodic, $r_{j+N} = r_j$, and the single cell dynamics is given by 
 \begin{equation}\label{2.5}
\frac{d}{dt} r_j=p_{j+1}-p_j\,,\qquad \frac{d}{dt}{p}_j=V'(r_j)-V'(r_{j-1})\,,
\end{equation}
$ j=1,\ldots,N$, together with the periodic boundary conditions
 $p_{1+N} = p_1$, $r_ 0= r_N$ and the constraint \eqref{2.3}. 
Through the stretch there is a coupling to the right neighbor and through the momentum a coupling to the left neighbor.
The potential is defined only up to translations, since the dynamics does not change under a simultaneous shift of $V(x)$ to $V(x-a)$ and $r_j$ to $r_j+a$, in other words, the potential can be shifted by shifting the initial $r$-field.
Note that our periodic boundary conditions are not identical to  fluid particles moving on a ring, but they may
become so for large system size when length fluctuations become negligible. Both equations are already of conservation type and we conclude that
\begin{equation}\label{2.7}
\frac{d}{dt}\sum^N_{j=1} r_j=0\,,\quad \frac{d}{dt}\sum^N_{j=1} p_j=0\,.
\end{equation}
We define the local energy by 
\begin{equation}\label{2.8}
e_j=\tfrac{1}{2}p^2_j + V(r_j)\,.
\end{equation}
Then its local conservation law reads
\begin{equation}\label{2.9}
\frac{d}{dt}e_j= p_{j+1} V'(r_j)-p_j V'(r_{j-1})\,,
\end{equation}
implying that
\begin{equation}\label{2.9a}
\frac{d}{dt}\sum^N_{j=1} e_j=0\,.
\end{equation}

At this point we assume that there are no further local conservation laws. Unfortunately our assumption, while reasonable, is extremely difficult to check. It certainly rules out the integrable chains,
which have $N$ conservation laws. There is no natural example known with, say, seven conservation laws. But the mere fact that there are
exceptions implies that close to integrability the predictions from fluctuating hydrodynamics could be on time scales which are not accessible. The parameters entering in
fluctuating hydrodynamics depend smoothly on the potential. Thus as one approaches, for example, the Toda potential 
no abrupt changes will be detected. In this sense, the theory cannot distinguish the Toda chain from a FPU chain both at
moderate temperatures. There is another limitation which can be addressed more quantitatively. If $V(x) +Px$ has a unique
minimum, then at very low temperatures the potential is close to a harmonic one. This feature will be properly reflected by fluctuating hydrodynamics through the temperature dependence of the coupling coefficients and the noise strength. 

The microcanonical equilibrium state is defined by the Lebesgue measure  constrained to a particular value of the conserved fields as
\begin{equation}\label{2.9b}
\sum^N_{j=1} r_j = \ell N \,,\quad \sum^N_{j=1} p_j=\mathsf{u}N\,,\quad \sum^N_{j=1} \big(\tfrac{1}{2}p^2_j+ V(r_j)\big) = \mathfrak{e} N
\end{equation}
with $\ell$ the stretch,  $\mathsf{u}$ the momentum, and $\mathfrak{e}$ the total energy per particle. In our context the equivalence of ensembles holds and computationally it is of advantage to switch to the canonical ensemble with respect to all three constraints. Then the dual variable for the stretch $\ell$ is the pressure $P$, for the momentum the average momentum,
again denoted by $\mathsf{u}$, and for the total energy $\mathfrak{e}$ the inverse temperature $\beta$. 
For the limit of infinite volume the  symmetric choice $j \in [-N,...,N]$ is more convenient. In the limit $N \to \infty$ either under the canonical equilibrium state, trivially, or under the microcanonical ensemble, by the equivalence of ensembles,  the collection $(r_j,p_j)_{j\in\mathbb{Z}}$ are independent random variables. Their single site probability density is given by
\begin{equation}\label{2.10}
Z(P,\beta)^{-1} \mathrm{e}^{-\beta(V(r_j)+Pr_j)} (2\pi/\beta)^{-1/2} \mathrm{e}^{-\frac{1}{2}\beta (p_j -\mathsf{u})^2}\,.
\end{equation}
Averages with respect to~(\ref{2.10}) are denoted by $\langle\cdot\rangle_{P,\beta,\mathsf{u}}$. The dependence on the 
average momentum can be removed by a Galilei transformation. Hence we mostly work with $\mathsf{u} = 0$, in which case we merely drop the index $\mathsf{u}$. We also introduce the internal energy, $\mathsf{e}$, through 
$\mathfrak{e} = \tfrac{1}{2}\mathsf{u}^2 +\mathsf{e}$, which agrees with the total energy at $\mathsf{u} = 0$. The canonical free energy, at $\mathsf{u} = 0$, is defined by
\begin{equation}\label{2.11}
G(P,\beta)=-\beta^{-1} \big(-\tfrac{1}{2}\log\beta + \log Z(P,\beta)\big)\,.
\end{equation}
Then
\begin{equation}\label{2.12}
\ell =\langle r_0\rangle_{P,\beta}\,,\quad \mathsf{e}=\partial_\beta\big(\beta G(P,\beta)\big) -P \ell=\frac{1}{2\beta}+\langle 
V(r_0)\rangle_{P,\beta}\,.
\end{equation}
The relation (\ref{2.12}) defines $(P,\beta) \mapsto (\ell(P,\beta),\mathsf{e}(P,\beta))$, thereby the inverse map 
$(\ell, \mathsf{e}) \mapsto (P(\ell,\mathsf{e}),$ $  \beta(\ell,\mathsf{e}))$, and thus accomplishes the switch between
the microcanonical thermodynamic variables $\ell, \mathsf{e}$ and the canonical thermodynamic variables $P, \beta$.

It is convenient to collect the conserved fields as the $3$-vector
$\vec{g} = (g_1,g_2,g_3)$, 
\begin{equation}\label{2.13}
\vec{g}(j,t) = \big(r_j(t),p_j(t),e_j(t)\big) \,,
\end{equation}
$\vec{g}(j,0) = \vec{g}(j)$. Then the conservation laws are combined as
\begin{equation}\label{2.14}
\frac{d}{dt}\vec{g}(j,t) + \vec{\mathcal{J}}(j+1,t)  - \vec{\mathcal{J}}(j,t)=0 
\end{equation}
with the local current functions 
\begin{equation}\label{2.15}
\vec{\mathcal{J}}(j) = \big( -p_j,-V'(r_{j-1}), - p_jV'(r_{j-1})\big)\,.
\end{equation}
Our prime interest are the equilibrium time correlations of the conserved fields, which are defined by
\begin{equation}\label{2.16}
S_{\alpha\alpha'}(j,t)=\langle g_{\alpha}(j,t) g_{\alpha'}(0,0)\rangle_{P,\beta} - \langle g_{\alpha}(0,0)\rangle_{P,\beta} \langle g_{\alpha'}(0,0)\rangle_{P,\beta}\,,
\end{equation}
$\alpha,\alpha'=1,2,3$. The infinite volume limit has been taken already and the average is with respect to thermal equilibrium at $\mathsf{u} = 0$. It is known that such a limit exists \cite{BeOl15}. Also the decay in $j$ is exponentially fast, but with a correlation length
increasing in time. Often it is convenient to regard $S(j,t)$, no indices, as a $3\times 3$ matrix. In general, $S(j,t)$ has certain symmetries, the first set resulting from space-time stationarity and the second set from time reversal, even for
$\alpha = 1,3$, odd for $\alpha = 2$, 
\begin{equation}\label{2.17}
S_{\alpha\alpha'}(j,t) = S_{\alpha'\alpha}(-j,-t)\,,\quad S_{\alpha\alpha'}(j,t) = (-1)^{\alpha+ \alpha'}S_{\alpha\alpha'}(j,-t) 
\,.
\end{equation}

At $t=0$ the average \eqref{2.16} reduces to a static average, which is easily computed.  The fields are uncorrelated in $j$, \textit{i.e.}
\begin{equation}\label{2.18}
S(j,0)=\delta_{j0} C
\end{equation}
with the static susceptibility matrix
\begin{equation}\label{2.19}
C=
\begin{pmatrix} \langle r_0;r_0\rangle_{P,\beta} & 0 & \langle r_0;V_0\rangle_{P,\beta} \\
                0 & \beta^{-1} & 0 \\
                \langle r_0;V_0\rangle_{P,\beta} & 0 & \frac{1}{2}\beta^{-2}+\langle V_0;V_0\rangle_{P,\beta}
\end{pmatrix}\,.
\end{equation}
Here, for $X$,$Y$ arbitrary random variables, $\langle X;Y\rangle=\langle XY\rangle-\langle X\rangle\langle Y\rangle$ denotes the second cumulant
and $V_0 = V(r_0)$, following the same notational convention as for $e_0$.
The conservation law implies the zeroth moment sum rule
\begin{equation}\label{2.21}
\sum_{j \in \mathbb{Z}}S_{\alpha\alpha'}(j,t) = \sum_{j \in \mathbb{Z}}S_{\alpha\alpha'}(j,0) = C_{\alpha\alpha'}\,.
\end{equation}

\begin{figure}[!ht]
\centering
\includegraphics[width=0.7\textwidth]{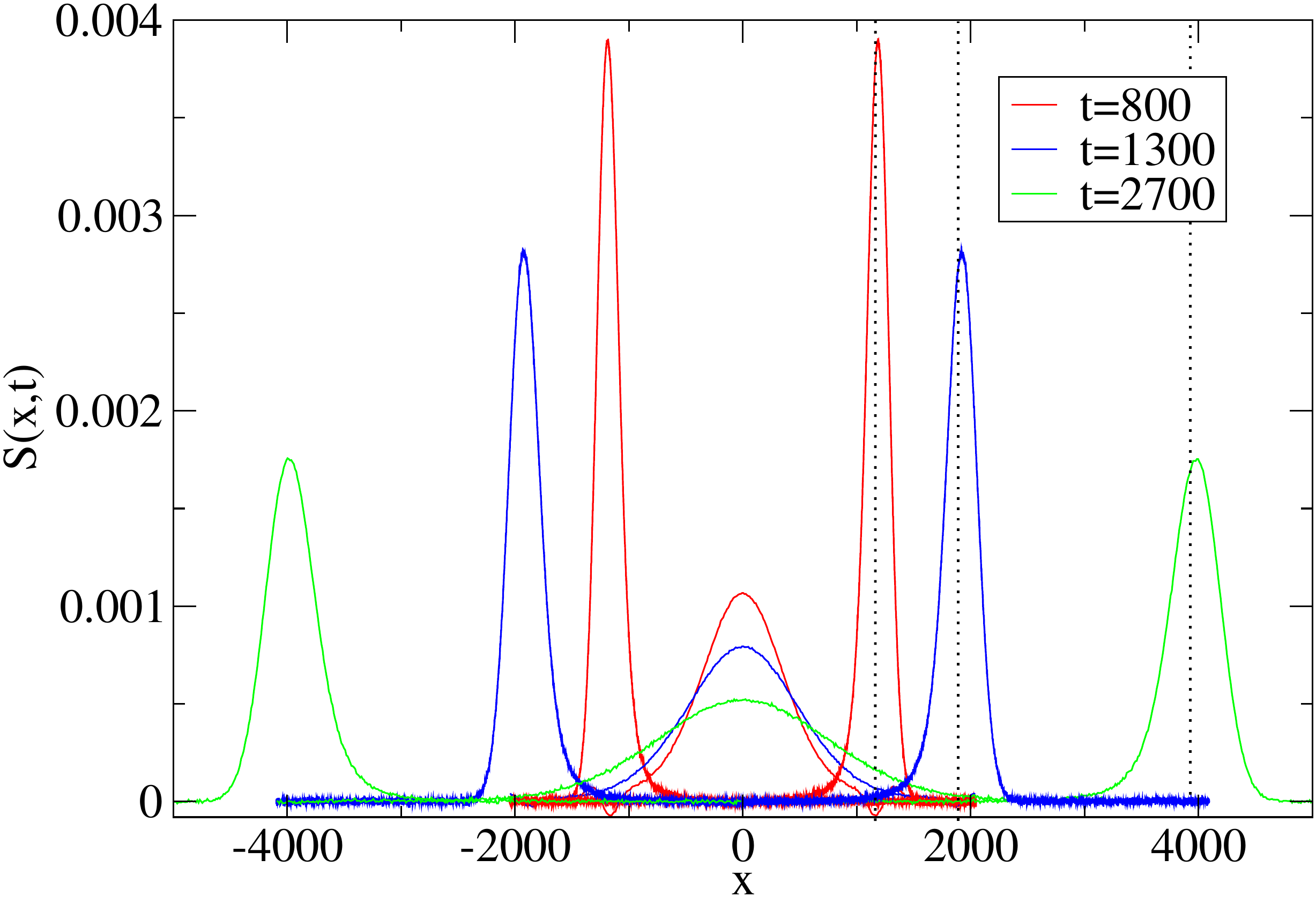}
\caption{Heat peak and sound peaks, area normalized to 1, at times $t = 800,1300,2700$, for a FPU chain with potential parameters $\alpha = 2$, $\beta = 1$,
pressure $P=1$, and temperature $\beta^{-1} = 0.5$.}
\label{figlabel0}
\end{figure}

An explicit computation of $S(j,t)$ is utterly out of reach. But with the current computer power MD simulations have become
an essential source of information. A broader coverage will be provided in Section \ref{sec5}. Just to have a first impression I show  in Fig. \ref{figlabel0} a recent MD simulation of the correlator for a FPU chain. One notes the central peak, called heat peak, which is standing still, and two
symmetrically located  peaks, called sound peaks which move outwards with the speed of sound. The peaks broaden with a certain power law which will have to be discussed. One expects, better hopes for, self-similar shape functions, at least for sufficiently long times. We do not know yet their form. But the central peak seems to have fat tails while the sound peaks fall off more rapidly, at least towards the outside of the sound cone. The area under each peak is preserved in time and normalized to 1 in our plot. If  the chain is initially perturbed near $0$ and  the response in one of the conserved fields is observed at $(j,t)$, then one records a signal, which is a linear combination of the peaks in Fig. \ref{figlabel0}, the coefficients depending on the initial perturbation. It might happen that one peak is missing. If the perturbation is orthogonal to 
all three physical fields, then there is no peak at all, only low amplitude noise.

Our goal is to predict the large scale structure of the correlator $S_{\alpha\alpha'}(j,t)$ encoding the propagation of local perturbations of the equilibrium state. On the crudest level, they should be  captured by linearized hydrodynamics, to which we turn next.\medskip\\
\textbf{Linearized hydrodynamics}. We start the dynamics from a product measure of the form \eqref{2.10}, but replace the uniform $ P, \mathsf{u},\beta$ by slowly varying spatial fields $P(\epsilon j),  \mathsf{u}(\epsilon j), \beta (\epsilon j)$ with $\epsilon \ll 1$. $\epsilon^{-1}$
is the macroscopic scale measured in lattice units. Equivalently we may regard $\epsilon $ as the lattice spacing. The relation \eqref{2.12} then defines also the slowly varying 
fields   $\ell(\epsilon j), \mathsf{u}(\epsilon j), \mathfrak{e}(\epsilon j)$. Because of the conservation laws the time change of such a state is slow and varies only over microscopic times of order $\epsilon ^{-1}t$ with macroscopic $t$ of order $1$.
We average Eq. \eqref{2.14} over the slowly varying initial state, which gives then the time change of the average locally conserved
fields. The expectation of the currents is more difficult. The difference in $j$ becomes $\epsilon\partial_x$ on the macroscopic
scale. Since the currents are functions depending only on at most two neighboring lattice sites, to lowest approximation their average can be computed in the equilibrium state with the corresponding 
local values of the fields, \textit{i.e.}  $\ell(\epsilon j, \epsilon^{-1} t), \mathsf{u}(\epsilon j,\epsilon^{-1} t), \mathfrak{e}(\epsilon j, \epsilon^{-1} t)$. Therefore we define the hydrodynamic Euler currents by the equilibrium averages
\begin{equation}\label{2.22}
\langle \vec{\mathcal{J}}(j)\rangle_{\ell,\mathsf{u},\mathfrak{e}} = \big(-\mathsf{u},P(\ell,\mathfrak{e}-\tfrac{1}{2}\mathsf{u}^2), \mathsf{u} P(\ell,\mathfrak{e}-\tfrac{1}{2}\mathsf{u}^2)\big) = \vec{\mathsf{j}}(\ell,\mathsf{u},\mathfrak{e})
\end{equation}
with $P(\ell,\mathsf{e})$ defined implicitly through~(\ref{2.12}). Our argument then leads to the macroscopic Euler equations 
\begin{equation}\label{2.22a}
\partial_t\ell -\partial_x \mathsf{u} =0\,,\quad
\partial_t \mathsf{u} +\partial_x P(\ell,\mathfrak{e}-\tfrac{1}{2}\mathsf{u}^2) =0\,,\quad
 \partial_t \mathfrak{e} +\partial_x \big( \mathsf{u} P(\ell,\mathfrak{e}-\tfrac{1}{2}\mathsf{u}^2)\big) =0\,.
\end{equation}
We refer to a forthcoming monograph \cite{BeOl15}, where the validity of the Euler equations is proved up to the first shock. Since, as emphasized already, it is difficult to deal with deterministic chaos, the authors add random velocity exchanges between neighboring particles which ensure that the dynamics locally enforces the microcanonical state.  

We are interested here only in small deviations from equilibrium and therefore linearize the Euler equations  as $\ell+u_1(x)$, $0+u_2(x)$, $\mathsf{e}+u_3(x)$ to linear order in the deviations $\vec{u}(x)$. This leads to the  linear equation
\begin{equation}\label{2.23}
\partial_t \vec{u}(x,t)+\partial_x A \vec{u} (x,t)=0
\end{equation}
with
\begin{equation}\label{2.24}
A=
\begin{pmatrix} 0 & -1 & 0 \\
               \partial_\ell P & 0 & \partial_\mathsf{e}P \\
                0 & P & 0
\end{pmatrix}\,.
\end{equation}
Here, and in the following, the dependence of $A$, $C$ and similar quantities on the background values $\ell,\mathsf{u} =0,\mathsf{e}$, hence on $P,\beta$,
 is suppressed from the notation. Beyond~(\ref{2.21}) there is the first moment sum rule which states that
\begin{equation}\label{2.25}
\sum_{j \in \mathbb{Z}}jS(j,t) = AC\,t\,.
\end{equation}
A proof, which in essence uses only the conservation laws and space-time stationarity of the correlations,
is given in \cite{Sp14}, see also see  \cite{To,Sc}. Microscopic properties enter only minimally. However, since $C = C^\mathrm{T}$ and $S(j,t)^\mathrm{T}
= S(-j,-t)$, Eq. \eqref{2.25} implies
the important relation
\begin{equation}\label{2.26}
  AC=(AC)^{\mathrm{T}}= CA^{\mathrm{T}}\,,
\end{equation}
with $^\mathrm{T}$ denoting transpose. Of course, \eqref{2.26} can be checked also directly from the definitions.
Since  $C>0$,  $A$ is guaranteed to have real eigenvalues
and a nondegenerate system of right and left eigenvectors. For $A$ one obtains the three eigenvalues $0, \pm c$
with 
\begin{equation}\label{2.27}
c^2= -\partial_{\ell} P+P \partial_\mathsf{e}P >0\,.
\end{equation}
Thus the solution to the linearized equation has three modes, one standing still, one right moving with velocity $c$ 
and one left moving with velocity $-c$. Hence we have identified the adiabatic sound speed as being equal to $c$.
 
\eqref{2.23} is a deterministic equation. But the initial data are random such that within our approximation
\begin{equation}\label{2.28}
\langle u_\alpha(x,0)u_{\alpha'}(x',0)\rangle = C_{\alpha\alpha'}\delta(x-x')\,.
\end{equation}
To determine the correlator $S(x,t)$ with such initial conditions is most easily achieved by introducing the linear transformation $R$ satisfying
\begin{equation}\label{2.29}
RAR^{-1} = \mathrm{diag}(-c,0,c)\,,\quad RCR^\mathrm{T} = 1\,.
\end{equation}
Up to trivial phase factors, $R$ is uniquely determined by these conditions. Explicit formulas are found in \cite{Sp14}.
Setting $\vec{\phi} = A \vec{u}$, one concludes
\begin{equation}\label{2.30}
\partial \phi_\alpha +c_\alpha \partial_x \phi_\alpha =0 \,,\quad \alpha= -1,0,1\,,
\end{equation}
with $\vec{c} = (-c,0,c)$. By construction, the random initial data have the correlator
\begin{equation}\label{2.31}
\langle \phi_\alpha(x,0)\phi_{\alpha'}(x',0)\rangle = \delta_{\alpha\alpha'} \delta(x-x')\,.
\end{equation}
Hence
\begin{equation}\label{2.32}
\langle \phi_\alpha(x,t)\phi_{\alpha'}(0,0)\rangle = \delta_{\alpha\alpha'}\delta(x - c_\alpha t)\,.
\end{equation}
We transform back to the physical fields. Then in the continuum approximation, at the linearized level,
\begin{equation}\label{2.33}
S(x,t) =R^{-1}\mathrm{diag}\big(\delta(x + ct),\delta(x ),\delta(x - c t)\big)R^{-\mathrm{T}}
\end{equation}
with $R^{-\mathrm{T}} = (R^{-1})^\mathrm{T}$.

Rather easily we have gained a crucial insight. $S(j,t)$ has three peaks which separate linearly in time.
For example, $S_{11}(j,t)$ has three sharp peaks moving with velocities $\pm c,0$.  Physically, one expects such peaks not to be strictly sharp, but to broaden in the course of time because of dissipation.  This issue will have to be explored in great detail. It follows from the zeroth moment sum rule that the area under each peak is preserved in time
and thus determined through \eqref{2.33}. Hence the   weights can be computed from the
matrix $R^{-1}$, usually called Landau-Plazcek ratios.  A Landau-Placzek ratio could  vanish,
either accidentally or by a particular symmetry.  An example is the momentum correlation $S_{22}(j,t)$. Since $ (R^{-1})_{20}=0$
always, its central peak is absent.

For integrable chains each conservation law generates a peak. Thus, \textit{e.g.}, $S_{11}(j,t)$ of the Toda chain is expected to have a broad spectrum expanding ballistically, rather than consisting of three sharp peaks.
\section{Nonlinear fluctuating hydrodynamics}\label{sec3}
\textbf{Euler currents to second order}. The broadening of the peaks results from random fluctuations in the currents, which tend to be uncorrelated in space-time.  Therefore the crudest model would be to assume that the current statistics  is space-time Gaussian white noise. In principle, the noise components could be correlated. But since the stretch current is itself conserved, its fluctuations will be taken care of by the momentum equation.
Momentum and energy currents have different signature under time reversal, hence their cross correlation vanishes.
As a result, there is a fluctuating momentum current of strength $\sigma_\mathsf{u}$ and an independent energy current 
of strength $\sigma_\mathsf{e}$. According to Onsager, noise is linked to dissipation as modeled by a diffusive term. Thus the linearized equations \eqref{2.23} are extended to 
\begin{equation}\label{3.1}
\partial_t \vec{\mathsf{u}}(x,t)+\partial_x \big( A \vec{\mathsf{u}} (x,t) - \partial_x  D\vec{\mathsf{u}}(x,t) + B\vec{\xi} (x,t)\big)=0\,.
\end{equation} 
Here $\vec{\xi} (x,t)$ is standard white noise with covariance
\begin{equation}\label{3.2}
\langle\xi_\alpha(x,t) \xi_{\alpha'}(x',t')\rangle= \delta_{\alpha\alpha'} \delta(x-x') \delta(t-t')
\end{equation}
and, as argued,  the noise strength matrix is diagonal as 
\begin{equation}\label{3.3}
B = \mathrm{diag}(0, \sigma_\mathsf{u},\sigma_\mathsf{e})\,.
\end{equation}
To distinguish the linearized Euler equations \eqref{2.23} from the Langevin equations \eqref{3.1}, we use 
$\vec{\mathsf{u}} = (\mathsf{u}_1,\mathsf{u}_2,\mathsf{u}_3)$ for the fluctuating fields.

From the introduction, we know already that a Gaussian fluctuation theory will fail. But still, it is useful to first explore the structure of the Langevin equation \eqref{3.1}.  The stationary measures for \eqref{3.1} are
spatial white noise with arbitrary mean. Since  small deviations from uniformity are considered, we always impose mean zero.  
Then  the components are correlated as
\begin{equation}\label{3.4}
\langle\mathsf{u}_\alpha(x) \mathsf{u}_{\alpha'}(x')\rangle= C_{\alpha\alpha'} \delta(x-x')\,.
\end{equation}
Stationarity relates the linear drift and the noise strength through the steady state covariance as
\begin{equation}\label{3.5}
- (AC -CA^\mathrm{T})\partial _x + (DC + CD^\mathrm{T})\partial_x^2 = BB^\mathrm{T} \partial_x^2\,.
\end{equation}
The first term vanishes by \eqref{2.26} and the diffusion matrix is uniquely determined as
\begin{equation}\label{3.6}
D=
\begin{pmatrix} 0 & 0 & 0 \\
              0 & D_\mathsf{u} & 0 \\
                
\tilde{D}_\mathsf{e}  & 0 & D_\mathsf{e}
\end{pmatrix}\,.
\end{equation}
with $\tilde{D}_\mathsf{e} = - \langle r_0;V_0\rangle_{P,\beta} \langle r_0;r_0\rangle_{P,\beta}^{-1}D_\mathsf{e}$. Here $D_\mathsf{u} >0$ is the momentum and $D_\mathsf{e}> 0 $ the energy diffusion coefficient, which are related to the noise strength as
\begin{equation}\label{3.7}
 \sigma_\mathsf{u}^2 = \langle p_0;p_0\rangle_{P,\beta} D_\mathsf{u} \,,\quad \sigma_\mathsf{e}^2 =   \langle e_0;e_0\rangle_{P,\beta}
 D_\mathsf{e}\,.
\end{equation}

We still have to establish that the stationary measure \eqref{3.4} is unique and is approached in the limit $t \to \infty$.
For this purpose it suffices that the $3\times 3$ matrix $ \mathrm{i}2\pi kA  - (2\pi k)^2D$ has its eigenvalues in the 
left hand complex plane, where for convenience we have switched to Fourier space with respect to $x$. If one drops $D$, then $\mathrm{i}2\pi kA$ has the eigenvalues $\mathrm{i}2\pi k c(-1,0,1)$. Hence one can use first order perturbation theory with respect to $-(2\pi k)^2D$, which is given by $\langle \tilde{\psi}_\alpha, D \psi_\alpha\rangle$, where $A\psi_\alpha = c_\alpha \psi_\alpha$ and $A^\mathrm{T}\tilde{\psi}_\alpha = c_\alpha \tilde{\psi}_\alpha$ are the right and left eigenvectors of $A$,
as listed in \cite{Sp14}. One simply has to follow the definitions and express every term through the cumulants of $r_0$ and $V_0$. As to be expected, the matrix elements from above have a definite value and the eigenvalues are shifted into the left hand complex plane. Similarly, $-D$ has eigenvalues $0,-D_\mathsf{u},-D_\mathsf{e}$ and the zero  eigenvalue is shifted to the left by second order  perturbation in $ \mathrm{i}(2\pi k)^{-1}A$.  The only condition is the strict positivity of $D_\mathsf{u}, D_\mathsf{e}$.

Based on \eqref{3.1} one computes the stationary space-time covariance, which most easily is written in Fourier space,
\begin{equation}\label{3.8}
S_{\alpha\alpha'}(x,t) = \langle \mathsf{u}_\alpha(x,t)  \mathsf{u}_{\alpha'}(0,0)\rangle  = \int dk\,\mathrm{e}^{\mathrm{i} 2 \pi
kx} \big(\mathrm{e}^{-\mathrm{i}t 2\pi kA - |t|(2\pi k)^2 D}C\big)_{\alpha\alpha'}\,. 
\end{equation}
To extract the long time behavior it is convenient to transform to normal modes. But before, we have to introduce
a more systematic notation. We will use the superscript $^\sharp$ for a normal mode quantity. Thus for the anharmonic chain
\begin{equation}\label{3.8a}
S^\sharp(j,t) = RS(j,t)R^\mathrm{T}\,,\quad S_{\alpha\alpha'}^\sharp(j,t)  = \langle (R\vec{g})_\alpha(j,t) (R\vec{g})_{\alpha'}(0,0)\rangle_{P,\beta}\,.
\end{equation}
The hydrodynamic fluctuation fields are defined on the continuum, thus functions of $x,t$, and we write
\begin{equation}\label{3.8b}
 S_{\alpha\alpha'}(x,t)  = \langle \mathsf{u}_\alpha(x,t)  \mathsf{u}_{\alpha'}(0,0)\rangle\,,\quad S^\sharp(x,t) = RS(x,t)R^\mathrm{T}\,.
\end{equation}
Correspondingly $A^\sharp = R A R^{-1} = \mathrm{diag}(-c,0,c)$, $D^\sharp= R D R^{-1}$, $B^\sharp = RB$.
Note that $\vec{\mathsf{u}}(x,t)$ will change its meaning when switching from linear to nonlinear fluctuating hydrodynamics.

In normal mode representation Eq. \eqref{3.8} becomes
\begin{equation}\label{3.9}
S^\sharp(x,t)   = \int dk\,\mathrm{e}^{\mathrm{i} 2 \pi
kx} \mathrm{e}^{-\mathrm{i}t 2\pi k A^\sharp- |t|(2\pi k)^2 D^\sharp}\,.
\end{equation}
The leading term, $\mathrm{i}t 2\pi kA^\sharp$, is diagonal, while the diffusion matrix $D^\sharp$ couples the components. 
But for large $t$ the peaks are far apart and the cross terms become small. More formally we split
$D^\sharp = D_\mathrm{dia} + D_\mathrm{off}$ and regard the off-diagonal part $D_\mathrm{off}$
as perturbation. When expanding, one notes that the off-diagonal terms carry an oscillating factor with frequency
$c_\alpha - c_\mathrm{\alpha'}$, $\alpha \neq \alpha'$. Hence these terms decay quickly and
 \begin{equation}\label{3.10}
S^\sharp_{\alpha\alpha'}(x,t)   \simeq \delta_{\alpha\alpha'} \int dk\,\mathrm{e}^{\mathrm{i} 2 \pi
kx} \mathrm{e}^{-\mathrm{i}t 2\pi k c_\alpha- |t|(2\pi k)^2 D^\sharp_{\alpha\alpha}}
\end{equation}
for large $t$. Each peak has a Gaussian shape function which broadens as $(D^\sharp_{\alpha\alpha}|t|)^{1/2}$.

Besides the peak structure, we have gained a second important insight. Since the peaks travel with distinct velocities,
on the linearized level the three-component system decouples into three scalar equations,
provided it is written in normal modes.

The linear fluctuation theory should be tested against adding nonlinear terms. The computation of the Introduction indicates that  in one dimension the quadratic part of the Euler current is a relevant perturbation. This can be seen even more directly by rescaling $\vec{\mathsf{u}} (x,t)$ to large space-time scales as $\vec{\mathsf{u}}_\epsilon (x,t) = 
\epsilon^{-b}\vec{\mathsf{u}}_\epsilon(\epsilon^{-1}x,\epsilon^{-z}t)$  and counting the powers of the nonlinear terms.
To have the correct $t=0$ covariance requires $b = 1/2$. The quadratic terms of the Euler currents are relevant, yielding
$z= 3/2$, and  the scaling exponents of cubic terms are only marginally relevant,
while a possible dependence of $D,B$ on $\vec{\mathsf{u}} (x,t)$ can be ignored.  
Thus, we retain the linear part \eqref{3.1} but expand the Euler currents 
including second order in $\vec{u}$, which turns \eqref{3.1} into the equations of nonlinear fluctuating hydrodynamics, \begin{eqnarray}\label{3.11}
&&\hspace{-20pt}\partial_t \mathsf{u}_1 -\partial_x \mathsf{u}_2 =0\,, \nonumber \\
&&\hspace{-20pt}\partial _t \mathsf{u}_2 +\partial_x\big((\partial_\ell P) \mathsf{u}_1 +(\partial_\mathsf{e}P) \mathsf{u}_3 + \tfrac{1}{2}
(\partial_\ell^2 P) \mathsf{u}_1^2 -\tfrac{1}{2} (\partial_\mathsf{e}P)\mathsf{u}_2^2 +\tfrac{1}{2}(\partial^2_\mathsf{e}P) \mathsf{u}_3^2 +
(\partial_\ell\partial_\mathsf{e}P) \mathsf{u}_1\mathsf{u}_3\nonumber\\
&&\hspace{260pt} - D_\mathsf{u}\partial_x \mathsf{u}_2 + \sigma_\mathsf{u}\xi_2\big) = 0\,,\nonumber\\
&&\hspace{-20pt}\partial _t \mathsf{u}_3 +\partial_x\big(P\mathsf{u}_2 +(\partial_\ell P) \mathsf{u}_1\mathsf{u}_2 +(\partial_\mathsf{e} P)\mathsf{u}_2\mathsf{u}_3 - \tilde{D}_\mathsf{e}
\partial_x \mathsf{u}_1 -D_\mathsf{e}\partial_x \mathsf{u}_3 +\sigma_\mathsf{e}\xi_3\big)=0\,.
\end{eqnarray} 
To explore their consequences is a more demanding task than solving the linear Langevin equation
and the results of the analysis will be more fragmentary.\medskip\\
\textbf{Stationary measure for the physical fields}. Adding quadratic terms could change drastically the character of the solution. 
To find out we first attempt 
to investigate the time-stationary measure. The vector field of the nonlinear part of \eqref{3.11} reads
 \begin{equation}\label{3.12}
\vec{F} = -\partial_x \big(0,  \tfrac{1}{2}
(\partial_\ell^2 P) \mathsf{u}_1^2 -\tfrac{1}{2} (\partial_\mathsf{e}P)\mathsf{u}_2^2 +\tfrac{1}{2}(\partial^2_\mathsf{e}P) \mathsf{u}_3^2 +(\partial_\ell\partial_\mathsf{e}P) \mathsf{u}_1\mathsf{u}_3,
(\partial_\ell P) \mathsf{u}_1\mathsf{u}_2 +(\partial_\mathsf{e} P)\mathsf{u}_2\mathsf{u}_3\big)\,.
\end{equation}
Formally the drift is divergence free, since
 \begin{equation}\label{3.13}
\sum_{\alpha = 1}^3\int dx\frac{\delta F_\alpha(x)}{\delta \mathsf{u}_\alpha(x)} = 
\partial_\mathsf{e}P\int dx(-\partial_x \mathsf{u}_2 + \partial_x \mathsf{u}_2) = 0
\end{equation}
and the infinite dimensional Lebesgue measure is invariant under the flow generated by $\vec{F}$. Since the equilibrium measure is a product, a natural ansatz for the invariant measure is 
Gaussian white noise retaining the physical susceptibility,
\begin{equation}\label{3.14}
\exp\Big( -\sum_{\alpha,\alpha'=1}^3 \tfrac{1}{2}(C^{-1})_{\alpha\alpha'}\int dx \mathsf{u}_\alpha(x)\mathsf{u}_{\alpha'}(x)\Big)
\prod_{\alpha,x}d\mathsf{u}_{\alpha}(x)\,.
\end{equation}
As established before, for the linear Langevin equation this measure is stationary. Thus, to find out whether it is also stationary for \eqref{3.11}, we only have to check the invariance
under the nonlinear flow
\begin{eqnarray}\label{3.15}
&&\hspace{-19pt}0= \frac{d}{dt} \sum_{\alpha,\alpha'=1}^3 (C^{-1})_{\alpha\alpha'}\int dx \mathsf{u}_\alpha(x)\mathsf{u}_{\alpha'}(x)
 = 2\sum_{\alpha=1}^3 \int dx(C^{-1} \vec{\mathsf{u}})_\alpha(x)\partial_t\mathsf{u}_{\alpha}(x)\nonumber\\
&&\hspace{-10pt} = \int dx a_0\mathsf{u}_2\partial_x\big((\partial_\ell^2P)\mathsf{u}_1^2 - (\partial_\mathsf{e}P)\mathsf{u}_2^2 +(\partial^2_\mathsf{e}P) \mathsf{u}_3^2 +2(\partial_\ell\partial_\mathsf{e} P) \mathsf{u}_1\mathsf{u}_3\big) \nonumber\\
&&\hspace{20pt}
+ 2\int dx(a_2\mathsf{u}_1 + a_3\mathsf{u}_3)\partial_x\big(
(\partial_\ell P) \mathsf{u}_1\mathsf{u}_2 +(\partial_\mathsf{e} P)\mathsf{u}_2\mathsf{u}_3\big)\,,
\end{eqnarray}
where $a_0 =(C^{-1})_{22}$,  $a_2=(C^{-1})_{13}$, $a_3=(C^{-1})_{33}$. The term cubic in $\mathsf{u}_2$ vanishes by the same argument as for the one-component case. All other terms are linear in $\mathsf{u}_2$, thus their sum has to vanish point-wise,
\begin{eqnarray}\label{3.16}
&&\hspace{-30pt}0= (a_0\partial_\ell^2 P - a_2\partial_\ell P)\partial_x \mathsf{u}_1^2 + (a_0\partial^2_\mathsf{e}P  -a_3
\partial_\mathsf{e}P)\partial_x \mathsf{u}_3^2 \nonumber\\
&&\hspace{10pt}+2\big(a_0(\partial_\ell\partial_\mathsf{e} P)\partial_x(\mathsf{u}_1\mathsf{u}_3) -a_2(\partial_\mathsf{e}P) \mathsf{u}_3\partial_x\mathsf{u}_1-a_3(\partial_\ell P) \mathsf{u}_1\partial_x\mathsf{u}_3\big)\,.
\end{eqnarray}
This leads to the constraints on the coefficients as
\begin{equation}\label{3.17}
a_0\partial_\ell^2 P = a_2\partial_\ell P\,,\quad a_0\partial^2_\mathsf{e}P  = a_3\partial_\mathsf{e} P\,,\quad 
a_0\partial_\ell\partial_\mathsf{e}P = a_3\partial_\ell P\,,\quad a_0\partial_\ell\partial_\mathsf{e} P = a_2\partial_\mathsf{e} P\,. 
\end{equation}
There are four constraints and five partial derivatives which may be regarded as independent parameters. Thus
one would expect that there is a sub-manifold in $V,P,\beta$ space for which \eqref{3.17} can be satisfied. We will come back to another representation of
these constraints below. Away from the special subset of  invariant Gaussian measures, we have no tools, but one would hope that the invariant measure still has a finite correlation length and exponential mixing.  Based on the mechanical model,
it is suggestive to assume $\mathsf{u}_2$ to be independent of $\mathsf{u}_1,\mathsf{u}_3$ and to have  white noise statistics. But this forces again the constraints 
\eqref{3.17} and results in the same Gaussian measure as before.

A basic property of the mechanical model is invariance under time reversal. On the level of fluctuating hydrodynamics this translates to the following property: We fix a time window $[0,T]$. Then, in the stationary process, the trajectories 
\begin{equation}\label{3.18}
(\mathsf{u}_1(t),  \mathsf{u}_2(t), \mathsf{u}_3(t))\,\,\mathrm{and}\,\, (\mathsf{u}_1(T-t), - \mathsf{u}_2(T-t), 
\mathsf{u}_3(T-t))
\end{equation}
with $0 \leq t \leq T$ have the same probability. To check \eqref{3.18} requires some information on the invariant measure. But under the Gaussian measure \eqref{3.14}, hence assuming the validity of the constraints, it can be shown that time reversal invariance indeed holds.\medskip\\ 
\textbf{Transformation to normal modes}. To  proceed further, it is convenient to write \eqref{3.11} in vector form,
 \begin{equation}\label{3.19a}
\partial_t \vec{\mathsf{u}}(x,t) +\partial_x \big(A \vec{\mathsf{u}}(x,t) +\tfrac{1}{2}\langle \vec{\mathsf{u}}, \vec{H} \vec{\mathsf{u}}\rangle -\partial_x \tilde{D} \vec{\mathsf{u}}(x,t) +\tilde{B}\vec{\xi}(x,t)\,\big)=0\,,
\end{equation}
where $\vec{H}$ is the vector consisting of the Hessians of the  currents with derivatives evaluated at the background values $(\ell,0,\mathsf{e})$,
\begin{equation}\label{3.19b}
  H^\alpha_{\gamma\gamma'} =\partial_{\mathsf{u}_\gamma} \partial_{\mathsf{u}_{\gamma'}} \mathsf{j}_\alpha\,,\qquad
 \langle \vec{\mathsf{u}}, \vec{H} \vec{\mathsf{u}}\rangle = \sum^3_{\gamma,\gamma'=1}\vec{H}_{\gamma\gamma'} \mathsf{u}_\gamma \mathsf{u}_{\gamma'}\,.
\end{equation}
As for the linear Langevin equation we transform to normal modes through 
 \begin{equation}\label{3.19}
 \vec{\phi} = R \vec{\mathsf{u}}\,.
\end{equation}
 Then 
 \begin{equation}\label{3.20}
\partial_t \phi_\alpha + \partial_x \big(c_\alpha \phi_\alpha + \langle\vec{\phi}, G^{\alpha}\vec{\phi}\rangle-\partial_x(D^\sharp\vec{\phi})_\alpha+(B^\sharp\vec{\xi})_\alpha\big)=0\,.
\end{equation}
By construction $B^\sharp B^{\sharp\mathrm{T}} = 2 D^\sharp $. The nonlinear coupling constants, denoted by $\vec{G}$,
are defined by
\begin{equation}\label{3.21}
 G^\alpha = \tfrac{1}{2}\sum^3_{\alpha'=1}  R_{\alpha\alpha'}  R^{-\mathrm{T}} H^{\alpha'}R^{-1}
\end{equation}
with the notation $R^{-\mathrm{T}} = (R^{-1})^\mathrm{T}$.

Since derived from a chain, the couplings are not completely arbitrary, but satisfy the symmetries 
\begin{eqnarray}\label{3.21a}
&&\hspace{-20pt}G^\alpha_{\beta\gamma} = G^\alpha_{\gamma\beta}\,, \quad G^\sigma_{\alpha\beta} 
= - G^{-\sigma}_{-\alpha-\beta}\,,\quad
 G^\sigma_{-10}  = G^\sigma_{01}  \,,  \nonumber\\
 &&\hspace{-20pt}G^0_{\sigma\sigma}  = - G^0_{-\sigma-\sigma}\,,\quad  
G^0_{\alpha\beta} = 0\,\,\,\, \mathrm{otherwise} \,.
\end{eqnarray}
In particular note that  
\begin{equation}\label{3.22}
 G_{00}^0 = 0\,,
 \end{equation} 
 always, while  $G^1_{11} = - G^{-1}_{-1-1}$ are generically different from 0. This property signals that the heat peak will behave differently from the sound peaks. The $\vec{G}$-couplings are listed in \cite{Sp14} and as a function of $P,\beta$ expressed in cumulants up to third order in $r_0,V_0$. The algebra is somewhat messy. But there is a short MATHEMATICA program available \cite{MeTUM} which, for given $P,\beta, V$, computes all coupling constants, including the matrices
 $C,A,R$. 
 
 We return to the issue of Gaussian time-stationary measures, where  we regard
 the coefficients $\vec{G},D^\sharp,B^\sharp$ as arbitrary, up to  $2D^\sharp = B^\sharp B^{\sharp\mathrm{T}}$, ignoring for a while their particular origin. 
The Langevin equation~(\ref{3.20}) is slightly formal. To have a well-defined evolution, we discretize space by a lattice of $N$ sites. The field $\vec{\phi}(x,t)$ then becomes $\vec{\phi}_j(t)$ with components $\phi_{j,\alpha}(t)$, $j=1,\ldots,N$, $\alpha=1,2,3$. The spatial  finite difference operator is denoted by $\partial_j$, $\partial_j f_j=f_{j+1}-f_{j}$, with transpose $\partial_j^\mathrm{T} f_j=f_{j-1}-f_{j}$. Then the discretized equations of fluctuating hydrodynamics read
\begin{equation}\label{B.1}
\partial_t \phi_{j,\alpha}+\partial_j\big(c_\alpha \phi_{j,\alpha} +\mathcal{N}_{j,\alpha} + \partial_j^\mathrm{T} D^\sharp\phi_{j,\alpha} + B^\sharp\xi_{j,\alpha}\big)=0
\end{equation}
with $\vec{\phi}_{j}=\vec{\phi}_{N+j}$, $\vec{\xi}_0=\vec{\xi}_N$, where $\xi_{j,\alpha}$ are independent Gaussian white noises with covariance
\begin{equation}\label{B.2}
\langle \xi_{j,\alpha}(t) \xi_{j',\alpha'} (t')\rangle =\delta_{jj'} \delta_{\alpha\alpha'} \delta(t-t')\,.
\end{equation}
The diffusion matrix $D^\sharp$ and noise strength $B^\sharp$ act on components, while the difference operator $\partial_j$ acts on the lattice site index $j$.

$\mathcal{N}_{j,\alpha}$ is quadratic in $\phi$. But let us first consider the case $\mathcal{N}_{j,\alpha} =0$. Then $\phi_{j,\alpha}(t)$ is a Gaussian process. The noise strength has been chosen such that one invariant measure is the Gaussian
\begin{equation}\label{B.3}
\prod^N_{j=1} \prod^3_{\alpha=1} \exp[-\tfrac{1}{2}\phi^2_{j,\alpha}] (2\pi)^{-1/2} d \phi_{j,\alpha}= \rho_\mathrm{G} (\phi) \prod^N_{j=1} \prod^3_{\alpha=1} d \phi_{j,\alpha}\,.
\end{equation}
Because of the conservation laws, the hyperplanes
\begin{equation}\label{B.4}
\sum^N_{j=1} \phi_{j,\alpha}=N\rho_\alpha\,,
\end{equation}
are invariant and on each hyperplane there is a Gaussian process with a unique invariant measure given by
 (\ref{B.3}) conditioned on that hyperplane.
For large $N$ it would become independent Gaussians with mean $\rho_\alpha$, our interest being the case of zero
mean,  $\rho_\alpha=0$. 

The generator of the diffusion process~(\ref{B.1}) with $\mathcal{N}_{j,\alpha}=0$ is given by
\begin{equation}\label{B.5}
L_0=\sum^N_{j=1} \Big(-\sum^3_{\alpha=1} \partial_j\big(c_\alpha \phi_{j,\alpha}+ \partial_j^\mathrm{T} D^\sharp\phi_{j,\alpha}\big) \partial_{\phi_{j,\alpha}}+\sum^3_{\alpha,\alpha'=1}  (B^\sharp B^{\sharp\mathrm{T}})_{\alpha\alpha'}  \partial_j \partial_{\phi_{j,\alpha}} \partial_j\partial_{\phi_{j,\alpha'}}\Big)\,.
\end{equation}
The invariance of $\rho_\mathrm{G}(\phi)$ can be checked through
\begin{equation}\label{B.7}
L^\ast_0 \rho_\mathrm{G}(\phi)=0\,,
\end{equation}
where $^\ast$ is the adjoint with respect to the flat volume measure. Furthermore linear functions evolve to linear functions according to
\begin{equation}\label{B.6}
\mathrm{e}^{L_0 t}\phi_{j,\alpha}= \sum^N_{j'=1} \sum^3_{\alpha'=1} (\mathrm{e}^{\mathcal{A}t})_{j\alpha,j'\alpha'} \phi_{j',\alpha'}\,,
\end{equation}
where the matrix $\mathcal{A}=- \partial_j\otimes \mathrm{diag} (c_1,c_2,c_3) - \partial_j \partial_j^\mathrm{T} \otimes D^\sharp$, the first factor acting on $j$ and the second on $\alpha$.

We now add the nonlinearity $\mathcal{N}_{j,\alpha}$. In general, this will modify the time-stationary measure and we have little control how. Therefore we propose to choose $\mathcal{N}_{j,\alpha}$ such that the corresponding vector field
$\partial_j\mathcal{N}_{j,\alpha}$ is divergence free \cite{SaSp09}. If $\mathcal{N}_{j,\alpha}$ depends only on the field at sites $j$ and $j+1$, then the unique solution reads
\begin{equation}\label{B.8}
\mathcal{N}_{j,\alpha}=\tfrac{1}{3}\sum^3_{\gamma,\gamma'=1} G^\alpha_{\gamma\gamma'}\big(\phi_{j,\gamma}\phi_{j,\gamma'}+\phi_{j,\gamma}\phi_{j+1,\gamma'}+\phi_{j+1,\gamma}\phi_{j+1,\gamma'}\big)\,.
\end{equation}
For $\rho_\mathrm{G}$ to be left invariant under 
the deterministic flow generated by the vector field $-\partial_j\mathcal{N}$ requires
\begin{equation}\label{B.9}
L_1\rho_\mathrm{G} = 0\,,\qquad L_1= - \sum^N_{j=1} \sum^3_{\alpha=1} \partial_j \mathcal{N}_{j,\alpha} \partial_{\phi_{j,\alpha}}\,,
\end{equation}
which implies
\begin{equation}\label{B.10}
\sum^N_{j=1} \sum^3_{\alpha=1} \phi_{j,\alpha} \partial_j \mathcal{N}_{j,\alpha}=0
\end{equation}
and thus the constraints
\begin{equation}\label{B.9a}
G^\alpha_{\beta\gamma}=G^\beta_{\alpha\gamma} \,\big(=G^\alpha_{\gamma\beta}\big)
\end{equation}
for all $\alpha,\beta,\gamma=1,2,3$, where in brackets we added the symmetry which holds by definition. Denoting the generator of the Langevin equation~(\ref{B.1}) by
\begin{equation}\label{B.10a}
  L=L_0+L_1\,,
\end{equation}
one concludes $L^\ast \rho_\mathrm{G}=0$, \textit{i.e.} the time-invariance of $\rho_\mathrm{G}$.

In the continuum limit the condition~(\ref{B.10}) reads
\begin{equation}\label{B.11}
\sum^3_{\alpha,\beta,\gamma=1} G^\alpha_{\beta\gamma} \int dx\phi_\alpha(x) \partial_x \big(\phi_\beta(x) \phi_\gamma(x)\big)=0\,,
\end{equation}
where $G^\alpha_{\beta\gamma}=G^\alpha_{\gamma\beta}$. By partial integration
\begin{equation}\label{B.13}
2 \sum^3_{\alpha,\beta,\gamma=1} G^\alpha_{\beta\gamma} \int dx \phi_\alpha(x) \phi_\beta(x) \partial_x \phi_\gamma(x)  
 =- \sum^3_{\alpha,\beta,\gamma=1} G^\alpha_{\beta\gamma} \int dx \phi_\beta(x) \phi_\gamma(x) \partial_x \phi_\alpha(x)\,.
\end{equation}
Hence (\ref{B.11}) is satisfied only if $G^\gamma_{\beta\alpha}=G^\alpha_{\beta\gamma}$, which is the condition 
(\ref{B.9a}) claimed for the discrete setting.

\eqref{B.9a} is the generalization of $\eqref{3.17}$, which is specific for the anharmonic chain. In fact, while abstractly true, it is not so easy to verify directly. But now we can  argue more convincingly why one should be allowed to
continue with assuming the validity of  the constraints \eqref{B.9a}. As we will discuss in the next section, the leading coupling constants are of the form 
$G^\alpha_{\alpha\alpha}$, while the sub-leading couplings have equal lower indices, $G^\alpha_{\gamma\gamma}$,
$\gamma \neq \alpha$. The off-diagonal matrix elements are irrelevant for the large scale behavior.
When one does the counting, all leading and sub-leading couplings can be chosen freely and the irrelevant couplings can be adjusted so that the constraint \eqref{B.9a} is satisfied. Appealing to universality,  the large space-time behavior should 
not depend on that particular choice. We expect that for general $\vec{G}$ the true time-stationary measure  will have short range correlations and nonlinear fluctuating hydrodynamics remains a valid approximation to the  dynamics of the anharmonic chain.

In related problem settings, a different point of view has been suggested \cite{SBM82,ZuMo83}. Firstly one notes that the Gaussian stationary measure \eqref{3.14}, hence also \eqref{B.3},
is simply inherited from the canonical equilibrium measure. In this respect there is no choice.
Also the nonlinear Euler currents are on the safe side. But the remaining terms are phenomenological to some extent. 
$D^\sharp,B^\sharp$ could depend on $\vec{\phi}$ itself. One could also include higher derivative terms.
In fact, one could try to choose the nonlinearities precisely in such a way that the dynamics is invariant under time-reversal and leaves the Gaussian measure invariant. The program as such may be easily endorsed, but so far I have not seen
 a convincing handling of the details.

\section{Mode-coupling theory}\label{sec4}
\textbf{Decoupling hypothesis}. For the linear equations the normal modes decouple for long times. The hypothesis
claims that such property persists when adding the quadratic nonlinearities. For the precise phrasing, we have to be somewhat careful. We consider a fixed component, $\alpha$, in normal mode representation. It travels with velocity $c_\alpha$, which is assumed to be distinct from all other mode velocities. If $G^\alpha_{\alpha\alpha} \neq 0$, then for the purpose of computing correlations of mode $\alpha$ at large scales, one can use the scalar conservation law
\begin{equation}\label{4.1}
\partial_t \phi_\alpha + \partial_x \big(c_\alpha \phi_\alpha + G^{\alpha}_{\alpha\alpha} \phi_\alpha^2-D^\sharp_{\alpha\alpha}\partial_x\phi_\alpha + B^\sharp_{\alpha\alpha}\xi_\alpha\big)=0\,,
\end{equation}
which coincides with the stochastic Burgers equation \eqref{1.18a}. If decoupling holds, one has the exact asymptotics as stated in \eqref{1.19} with $\lambda = 2\sqrt{2}|G^\alpha_{\alpha\alpha}|$. The universal scaling function $f_{\mathrm{KPZ}}$ is tabulated in~\cite{Prhp}, denoted there by $f$. $f_{\mathrm{KPZ}}\geq 0$, $\int dxf_{\mathrm{KPZ}}(x)=1$, $f_{\mathrm{KPZ}}(x)=f_{\mathrm{KPZ}}(-x)$, $\int dxf_{\mathrm{KPZ}}(x)x^2 \simeq 0.510523$. $f_{\mathrm{KPZ}}$ looks like a Gaussian with a large $|x|$ decay as $\exp[-0.295|x|^{3}]$. Plots are provided in~\cite{PrSp04,Prhp}.

For an anharmonic chain, $G_{00}^0
=0$ always and the decoupling hypothesis applies only to the sound peaks, provided $G^1_{11} = -G^{-1}_{-1-1}\neq 0$
which generically is the case. If $G^1_{11}\neq 0$, then the \textit{exact} scaling form is
\begin{equation}\label{4.2}
S^{\sharp}_{\sigma\sigma}(x,t)\cong (\lambda_\mathrm{s} t)^{-2/3} f_{\mathrm{KPZ}}
\big((\lambda_\mathrm{s} t)^{-2/3}(x - \sigma c t)\big)\,,\quad \lambda_\mathrm{s} = 2 \sqrt{2} |G^\sigma_{\sigma\sigma}|\,,
 \end{equation}
$\sigma = \pm 1$. To find out about the scaling behavior of the heat mode other methods have to be developed. 

For one-dimensional fluids, van Beijeren \cite{vB13} follows the scheme developed in \cite{ErHa76}  and arrived first at the prediction \eqref{4.2}
together with the L\'{e}vy 5/3 heat peak to be discussed below. 
In  \cite{vB13} no Langevin equations appear. I regard them as a useful intermediate step valid on a mesoscopic scale. In the Langevin form the theory can be applied to a large class of one-dimensional systems. As a tool, fluctuating hydrodynamics has been proposed considerably earlier \cite{NaRa02} and used to predict the $t^{-2/3}$ decay of the total energy current correlation.
\medskip\\
\textbf{One-loop, diagonal, and small overlap approximations}. 
We return to the Langevin equation (\ref{B.1}) and consider the mean zero, stationary $\phi_{j,\alpha}(t)$ process
with $\rho_\mathrm{G}$ as $t=0$ measure.  The stationary covariance reads
\begin{equation}\label{B.14}
S^\sharp_{\alpha\alpha'}(j,t)=\langle \phi_{j,\alpha}(t)\phi_{0,\alpha'}(0)\rangle =\langle \phi_{0,\alpha'}\mathrm{e}^{Lt}\phi_{j,\alpha}\rangle_{\mathrm{eq}}\,,\quad t \geq 0\,.
\end{equation}
On the left, $\langle\cdot\rangle$ denotes the average with respect to the stationary $\phi_{j,\alpha}(t)$ process and on the right 
$\langle\cdot\rangle_{\mathrm{eq}}$ refers to the average with respect to $\rho_\mathrm{G}$.
By construction
\begin{equation}\label{B.15}
S^\sharp_{\alpha\alpha'}(j,0)=\delta_{\alpha\alpha'} \delta_{j0}\,.
\end{equation}
The time derivative reads
\begin{equation}\label{B.16}
\frac{d}{dt} S^\sharp_{\alpha\alpha'}(j,t)=\langle \phi_{0,\alpha'}(\mathrm{e}^{Lt}L_0 \phi_{j,\alpha})\rangle_{\mathrm{eq}}+ \langle \phi_{0,\alpha'}(\mathrm{e}^{Lt}L_1 \phi_{j,\alpha})\rangle_{\mathrm{eq}}\,.
\end{equation}
We insert
\begin{equation}\label{B.17}
\mathrm{e}^{Lt}=\mathrm{e}^{L_0 t}+\int^t_0 ds\, \mathrm{e}^{L_0(t-s)} L_1 \mathrm{e}^{Ls}
\end{equation}
in the second summand of~(\ref{B.16}). The term containing only $\mathrm{e}^{L_0 t}$ is cubic in the time zero fields and hence its average vanishes.  Therefore one arrives at 
\begin{equation}\label{B.18}
\frac{d}{dt} S^\sharp_{\alpha\alpha'}(j,t)=  \mathcal{A} S_{\alpha\alpha'}(j,t) + \int^t_0 ds \langle \phi_{0,\alpha'} \mathrm{e}^{L_0(t-s)}L_1(\mathrm{e}^{Ls} L_1 \phi_{j,\alpha}) \rangle_\mathrm{eq}\,.
\end{equation}
For the adjoint of $\mathrm{e}^{L_0(t-s)}$ we use~(\ref{B.6}) and for the adjoint of $L_1$ we use
\begin{equation}\label{B.18a}
\langle \phi_{j,\alpha}  L_1  F(\phi)\rangle_\mathrm{eq} =  - \langle (L_1 \phi_{j,\alpha})   F(\phi)\rangle_\mathrm{eq} \,,
\end{equation}
which both rely on $\langle \cdot \rangle_\mathrm{eq}$ being the average with respect to $\rho_\mathrm{G}$. Furthermore
\begin{equation}\label{B.19}
L_1 \phi_{j,\alpha} = - \partial_j \mathcal{N}_{j,\alpha}\,.
\end{equation}
 Inserting in~(\ref{B.18}) one arrives at the identity
\begin{equation}\label{B.18b}
 \frac{d}{dt} S^\sharp_{\alpha\alpha'}(j,t)= \mathcal{A}S^\sharp_{\alpha\alpha'}(j,t)
  - \int^t_0 ds  \langle
(\mathrm{e}^{\mathcal{A}^\mathrm{T}(t-s)}\partial_j \mathcal{N}_{0,\alpha'})(\mathrm{e}^{Ls} \partial_j \mathcal{N}_{j,\alpha}) \rangle_\mathrm{eq}\,.
\end{equation}

To obtain a closed equation for $S^\sharp$ we note that the average 
\begin{equation}\label{B.19a}
\langle \partial_{j'}  \mathcal{N}_{j',\alpha'} \mathrm{e}^{L_s} \partial_j 
\mathcal{N}_{j,\alpha}\rangle_\mathrm{eq} = \langle \partial_{j } \mathcal{N}_{j,\alpha}(s)\partial_{j'} 
\mathcal{N}_{j',\alpha'}(0)\rangle
\end{equation}
is
a four-point correlation. We invoke the Gaussian factorization as
\begin{equation}\label{B.20a}
\langle\phi(s)\phi(s)\phi(0)\phi(0)\rangle\cong \langle\phi(s)\phi(s)\rangle\langle\phi(0)\phi(0)\rangle+2\langle\phi(s)\phi(0)\rangle\langle\phi(s)\phi(0)\rangle\,.
\end{equation}
 The first summand vanishes because of the difference operator $\partial_j$. Secondly we replace the bare propagator $\mathrm{e}^{\mathcal{A}(t-s)}$ by the interacting propagator $S^\sharp(t-s)$, which corresponds to a partial resummation of the perturbation series in $\vec{G}$. Finally we take a limit of zero lattice spacing. This step could be avoided, and is done so in our numerical scheme for the mode-coupling equations. We could also maintain the ring geometry which,
  for example,  would allow 
 to investigate  collisions between the moving peaks. Universality is only expected for large $j,t$, hence in the limit of zero lattice spacing. The continuum limit of $S^\sharp(j,t)$ is denoted by $S^\sharp(x,t)$, $x\in\mathbb{R}$. With these steps we arrive at the mode-coupling equation
 \begin{eqnarray}\label{B.20}
&&\hspace{-53pt}  \partial_t S^\sharp_{\alpha\beta}(x,t)= \sum^3_{\alpha'=1} \Big(\big(-c_\alpha\delta_{\alpha\alpha'}\partial_x +D_{\alpha\alpha'}\partial^2_x\big) S^\sharp_{\alpha'\beta}(x,t)\nonumber\\
&&\hspace{45pt} + \int^t_0 ds \int_{\mathbb{R}} dy    M_{\alpha\alpha'}(y,s) \partial^2_xS^\sharp_{\alpha'\beta}(x-y,t-s)\Big)
\end{eqnarray}
with the memory kernel
\begin{equation}\label{B.21}
M_{\alpha\alpha'}(x,t)= 2\sum^3_{\beta',\beta'',\gamma',\gamma''=1} G^\alpha_{\beta'\gamma'} G^{\alpha'}_{\beta''\gamma''} S^\sharp_{\beta'\beta''}(x,t) S^\sharp_{\gamma'\gamma''}(x,t)\,.
\end{equation}

In numerical simulations of both, the mechanical model of anharmonic chains and the mode-coupling equations,  it is consistently observed that $S^{\sharp}_{\alpha\alpha'}(j,t)$ becomes  approximately  diagonal fairly rapidly. To analyse the long time asymptotics on the basis of \eqref{B.20} we therefore rely on the diagonal approximation
\begin{equation}\label{40}
  S^{\sharp}_{\alpha\alpha'}(x,t)\simeq \delta_{\alpha\alpha'} f_\alpha(x,t)\,.
\end{equation}
Then $f_\alpha(x,0)=\delta(x)$ and the $f_\alpha$'s satisfy
\begin{equation}\label{41}
\partial_t f_\alpha(x,t)= (-c_\alpha \partial_x+D^\sharp_{\alpha\alpha} \partial^2_x) f_\alpha (x,t) + \int^t_0 ds \int_{\mathbb{R}} dy
 \partial^2_x  f_\alpha(x-y,t-s) M_{\alpha\alpha}(y,s)\,,
\end{equation}
$\alpha=-1,0,1$,  with memory kernel
\begin{equation}\label{41a}
M_{\alpha\alpha}(x,t)=2 \sum_{\gamma,\gamma'=0,\pm 1} (G^\alpha_{\gamma\gamma'})^2 f_\gamma(x,t) f_{\gamma'}(x,t)\,.
\end{equation}

The solution to \eqref{41} has two sound peaks centered at 
$\pm ct$ and the heat peak  sitting at $0$. All three peaks  have a width much less than $ct$. But then,
in case $\gamma\neq \gamma'$, the product  $f_\gamma(x,t) f_{\gamma'}(x,t) \simeq 0$ for large $t$. Hence for the memory kernel \eqref{41a} we invoke a small overlap approximation as
\begin{equation}\label{43}
M_{\alpha\alpha}(x,t)\simeq M^{\mathrm{dg}}_{\alpha}(x,t)=2\sum_{\gamma=0,\pm 1}(G^\alpha_{\gamma\gamma})^2 f_\gamma(x,t)^2\,,
\end{equation}
which is to be inserted in Eq. \eqref{41}. \medskip\\
\textbf{Numerical simulations of the mode-coupling equations}. When starting this project together with Christian Mendl, in the summer of 2012 we spent  
many days in numerically simulating the mode coupling equations with initial conditions 
$S^\sharp_{\alpha\alpha'}(j,0) = \delta_{\alpha\alpha'}\delta_{0j}$.  Only a few plots are in print \cite{MeSp13}, simply because there is such a large parameter space and it is not clear where to start and where to end. Still, for our own understanding this period was extremely helpful. Mostly we simulated in Fourier space. System size was up to $400$. Speeds were of order $1$, thereby limiting the simulation time to about 200,
the time of the first peak collision. For such sizes the simulations are fast and many variations could be explored.
We started from the scalar equation, to be discussed below, moved up to two modes, and eventually to three modes
with parameters taken from an actual anharmonic chain. $|\vec{G}^\alpha_{\alpha'\alpha'}|$ was 
either $0$ or somewhere in the range $0.3$ to $2.5$. $D^\sharp$ is a free parameter which was varied from $0$ to 
$|\vec{G}^\alpha_{\alpha'\alpha'}|/2$. We always simulated the complete matrix-valued mode-coupling equations
\eqref{B.20}. Our main findings can be summarized as:\\
\textit{(i)} For a large range of parameters, the diagonal approximation in generally failed for short times, but was quickly restored with the off-diagonal elements being at most $10\%$ of the diagonal ones.\\
\textit{(ii)} The results were  fairly insensitive to the choice of $D^\sharp$. In fact, $D^\sharp=0$ works also well. Apparently the memory term generates already enough dissipation.\\
\textit{(iii)} We varied the overlap coefficients $G^\alpha_{\gamma\gamma'}$ with $\gamma\neq\gamma'$.
Over the time scale of the simulation no substantial changes were observed.

All these findings confirm the approximations proposed.

As our biggest surprise, except for trivial cases we never reached the asymptotic regime. The peak structure develops fairly rapidly. The peak shape then changes slowly, roughly consistent with the predicted scaling exponents, but it does not reach a self-similar form. For example in the case of the sound peak, on the scale $t^{2/3}$, rather than being symmetric, as claimed by \eqref{4.2}, it is still badly distorted, tilted away from the central peak with rapid decay outside the sound cone but rather slow power law type of decay towards the central peak. To improve one would have to simulate larger system sizes and longer times. But then  numerical simulations become heavy and the fun evaporates. More attention can be achieved by molecular dynamics (MD) simulation of the mechanical  chain.

For given parameters $V,P,\beta$ one easily computes all the required coefficients. So one goal was to run a MD and put the results on top of the ones from a simulation
of the mode coupling equations. For this to be a reasonable program, one would have to simulate the mode-coupling equations  for sizes of $N = 4000$ and more,
which we never attempted.

For the scalar case the situation is much simpler. The mode-coupling equation takes the form 
\begin{equation}\label{44}
\partial_t f(x,t)= D \partial^2_x f(x,t)+ 2G^2 \int^t_0 ds \int_{\mathbb{R}} d y
 \partial^2_xf(x-y,t-s) f(y,s)^2\,,
\end{equation}
which is the one-loop approximation for the stochastic Burgers equation  \cite{vBKS85} . For large $x,t$, its solution with 
initial condition $f(x,0) = \delta(x)$ takes the scaling form
\begin{equation}\label{45}
f(x,t)\cong (\lambda_\mathrm{s} t)^{-2/3} f_{\mathrm{mc}} \big((\lambda_\mathrm{s} t)^{-2/3}x\big)\,.
\end{equation}
Inserting in \eqref{44}, one first finds the
non-universal scaling coefficient
\begin{equation}\label{46}
\lambda_\mathrm{s} = 2\sqrt{2}|G^\sigma_{\sigma\sigma}|\,.
\end{equation}
Secondly $\hat{f}_{\mathrm{mc}}$, the Fourier transform of  ${f}_{\mathrm{mc}}$, is defined as solution of the fixed point equation
\begin{equation}\label{45a}
\tfrac{2}{3} \hat{f}'_{\mathrm{mc}}(w) = - \pi^2 w \int_0^1 ds \hat{f}_{\mathrm{mc}}((1-s)^{2/3}w) \int_{\mathbb{R}} dq
\hat{f}_{\mathrm{mc}}(s^{2/3}(w-q)) \hat{f}_{\mathrm{mc}}(s^{2/3} q)
\end{equation}
with $w \geq 0$ and $\hat{f}_{\mathrm{mc}}(0) = 1$, $\hat{f}'_{\mathrm{mc}}(0) = 0$.

Eq. \eqref{45a} is based on the closure assumption \eqref{B.20a} and there is no reason to infer that it is exact. However from 
our numerical simulations we conclude that $f_{\mathrm{KPZ}}$ differs from $f_{\mathrm{mc}}$ by a few percent only.
We regard the scalar case as a strong support for the entire approach. But for several components the large finite size effects
prohibit one to arrive at a similarly simple claim.\medskip\\
\textbf{Asymptotic self-similarity}. Within mode-coupling the asymptotic
shape function for the sound peaks is given by 
\begin{equation}\label{45b}
f_\sigma(x,t)\cong (\lambda_\mathrm{s} t)^{-2/3} f_{\mathrm{mc}} \big((\lambda_\mathrm{s} t)^{-2/3}(x- \sigma ct)\big)\,,
\end{equation}
$\sigma = \pm 1$.
For the heat peak we employ \eqref{41} together with
\eqref{43}, using as an input that  the asymptotic form of $f_\sigma$ is known already. In fact the scaling exponent for 
$f_\sigma$ is crucial, but the precise shape of $f_\sigma$ enters only mildly. Hence, again switching to Fourier space, one has to solve
\begin{eqnarray}\label{47}
&&\hspace{-30pt}\partial_t \hat{f}_0(k,t)= -D^\sharp_0 (2\pi k)^2  \hat{f}_0(k,t)\nonumber\\
&&\hspace{0pt} -2\sum_{\sigma=\pm 1} (G^0_{\sigma\sigma})^2(2\pi  k)^2 \int^t_0 ds  \hat{f}_0(k,t-s) 
\int_{\mathbb{R}}  dq  \hat{f}_\sigma(k-q,s)  \hat{f}_\sigma(q,s)\,,
\end{eqnarray}
$ \hat{f}_0(k,0)=1$. For $\hat{f}_\sigma$ one inserts the asymptotic result~(\ref{45}). (\ref{47}) is a linear equation which is solved through Laplace transform with the result
\begin{equation}\label{48}
\hat{f}_0(k,t)\cong \mathrm{e}^{-|k|^{5/3} \lambda_\mathrm{h} t}\,,
\end{equation}
where
\begin{eqnarray}\label{49}
&&\hspace{-15pt}\lambda_\mathrm{h}= \lambda^{-2/3}_\mathrm{s} (G^0_{\sigma\sigma})^2 (4 \pi)^2  \int^\infty_0 dt t^{-2/3} \,\cos (2 \pi ct) \int_{\mathbb{R}}  dx f_{\mathrm{mc}}(x)^2 \nonumber\\
&&\hspace{0pt} =  \lambda^{-2/3}_\mathrm{s} (G^0_{\sigma\sigma})^2 (4 \pi)^2 (2\pi c)^{-1/3}
\tfrac{1}{2}\pi \frac{1}{\Gamma(\tfrac{2}{3})}\frac{1}{ \cos(\tfrac{\pi}{3})} \int_{\mathbb{R}}  dx f_{\mathrm{mc}}(x)^2
\end{eqnarray} 
and we used the symmetry $G^0_{\sigma\sigma} = -G^0_{-\sigma-\sigma}$, see \cite{Sp14} for details. (\ref{48}) is the Fourier transform of the symmetric $\alpha$-stable distribution with exponent $\alpha=5/3$, also known as L\'{e}vy distribution. In real space the asymptotics reads, for $|x|\geq (\lambda_\mathrm{h} t)^{3/5}$,
\begin{equation}\label{50}
f_0(x,t)\simeq \pi^{-1} \lambda_\mathrm{h} t |x|^{-8/3}\,.
\end{equation}
$f_\mathrm{mc}$ is a smooth function with rapid decay. On the other hand, $f_0$ has fat tails and its variance is divergent. According to~(\ref{50}), at $x=\sigma ct$ the heat peak $f_0(\sigma ct,t)\cong \pi^{-1} \lambda_\mathrm{h} c^{-8/3} t^{-5/3}$. This explains why there is still coupling between $f_0$ and $f_\sigma$, despite the large spatial separation. In fact, numerically one observes that beyond the sound cone, $x = \pm ct$, the solution decays exponentially fast. As $t$ becomes large the tails of $f_0$ are build up between the two sound peaks, so to speak they unveil the 
L\'{e}vy distribution. 

In \eqref{47} we could also insert for $f_\sigma$ the exact scaling function $f_{\mathrm{KPZ}}$, which would slightly modify  
$\lambda_\mathrm{h}$. $f_{{\rm L\acute{e}vy}}$ is an approximation, just as $f_{\mathrm{mc}}$. But the MD simulations display so convincingly the L\'{e}vy distribution that one might be willing to regard it as exact. If so, the exact 
$\lambda_\mathrm{h}$ must be based on $f_{\mathrm{KPZ}}$. To obtain the correlations of the physical fields, one has to use
\begin{equation}\label{51}
S(j,t) = R^{-1}S^\sharp(j,t)R^{-\mathrm{T}}\,.
\end{equation} 
In particular the correlations of the physical fields are given through
\begin{equation}\label{51a}
S_{\alpha\alpha}(j,t) = \sum_{\sigma = 0,\pm{1}}|(R^{-1})_{\alpha\sigma}|^2 f_{\sigma}(j,t)\,,
\end{equation} 
where for $f_\sigma$ the asymptotic scaling form is inserted.
Then asymptotically the $\ell$-$\ell$ and the $\mathsf{e}$-$\mathsf{e}$ correlations show generically all three peaks. However, for
the $\mathsf{u}$-$\mathsf{u}$ correlations the central peak is  missing asymptotically, since  $(R^{-1})_{20} = 0$ independently of the interaction potential $V$.

We note that the coefficient $D^\sharp$ does not appear in the asymptotic scaling form, of course neither $B^\sharp 
B^{\sharp\mathrm{T}}= 2 D^\sharp$. This result is consistent with the picture that noise and dissipation are required to maintain the correct local
stationary measure with susceptibility $C$. The long time asymptotics is however governed by the nonlinearities. 
\medskip\\
\textbf{No signal beyond the sound cone}. Physically the sound speed is an upper limit for the propagation of
small disturbances. Since the initial state has a finite correlation length, one would expect that towards the outside of the sound cone correlations decay exponentially, while inside the sound cone there seems to be no particular restriction.
As a consistency check, one would hope that such a general feature is properly reproduced by mode-coupling. Their
numerical solutions conform with this expectation, at least for the small system sizes explored. But for the scaling limit one has to let $t \to \infty$ and the decay information seems to be lost. However there is still a  somewhat subtle trace.

To explain, we have to first recall some properties of  L\'{e}vy stable distributions. Except for trivial rescalings, they are characterized by two parameters, traditionally called $\alpha,\beta$, where for simplicity we momentarily stick to this convention, without too much risk of confusion.
The probability density has a simple form in Fourier space,
\begin{equation}\label{4.3}
\hat{f}_{{\rm L\acute{e}vy},\alpha,\beta}(k) = \exp\big(-|k|^\alpha \big[1-\mathrm{i} \beta \tan(\tfrac{1}{2}\pi \alpha) \mathrm{sgn}(k)\big]\big).
\end{equation}
The parameter $\alpha$  controls the steepness, $0 < \alpha < 2$, while $\beta$  controls the asymmetry, 
$|\beta| \leq 1$. For $|\beta| > 1$ the Fourier integral no longer defines a non-negative function. At the singular point $\alpha = 2$ only $\beta = 0$ is admitted and the probability density is  a Gaussian.  If $|\beta|< 1$, the asymptotic decay of  $f_{{\rm L\acute{e}vy},\alpha,\beta}(x)$  is determined by $\alpha$ and is given by $|x|^{-\alpha-1}$ for $|x| \to \infty$. At $|\beta| = 1$ the two tails show different decay. The functions corresponding to $\beta = 1$ and $\beta = -1$ are mirror images, for $\beta=1$ the slow decay being for $x \to -\infty$ and still as $|x|^{-\alpha-1}$. For $0 < \alpha \leq 1$, $f_{{\rm L\acute{e}vy},\alpha,1}(x) = 0$ for $x > 0$, while for $1 < \alpha < 2$ the decay becomes stretched exponential as
$\exp(-c_0 x^{\alpha/(1 - \alpha)})$ with known constant $c_0$. We refer to~\cite{UZ99} for more details.

For the heat peak we obtained the symmetric  L\'{e}vy distribution because the sound peaks are reflection symmetric,
implying $c_1 = -c_{-1}$ and $(G^0_{11})^2 = (G^0_{-1-1})^2$. If hypothetically we would choose distinct couplings, or $c_1 \neq - c_{-1}$,
then this imbalance would produce a $\beta \neq 0$. If one of the sound peaks would be completely missing, as in the case for a system with only two conserved fields, then necessarily
$|\beta| = 1$. In accordance with the physical principle, the sign of $\beta$ is  such that the fast decay of  $f_{{\rm L\acute{e}vy},\alpha,\pm 1}(x)$  is towards the outside of the sound cone, while the slow decay is towards the single sound peak. For finite $t$, this slow decay
will be cut by the sound peak. Thus the scaling solution of the mode-coupling equations reproduces the rapid decay towards the exterior of the sound cone. This is a completely general fact, any number of components and any $\vec{G}$
\cite{Sc15}.\medskip\\ 
\textbf{Dynamical phase diagram}. As already indicated through the particular case $G^0_{00} = 0$, the large scale structure of the solution depends on whether $G^\alpha_{\gamma\gamma} = 0$ or not. One extreme case would be $G^{\alpha}_{\alpha\alpha} \neq 0$ for all
$\alpha$, implying that the three peaks have KPZ scaling behavior. The other extreme is $G^{\alpha}_{\gamma\gamma} = 0$
for all  $\alpha,\gamma$, resulting in all peaks to have diffusive broadening. For the case of only  two modes, the full phase diagram has seven distinct phases, with unexpected details worked out in \cite{StSp14}. For the general case of $n$ components the long time asymptotics is completely classified in \cite{PSSS15,Sc15}. Anharmonic chains have special symmetries and not all possible couplings $\vec{G}$ can be realized. 
Given that $G^0_{00} = 0$ and because the sound peaks are symmetric, to have a distinct scaling requires
\begin{equation}\label{4.4}
G^1_{11} =0\,,
\end{equation}
which can be realized. The behavior is then determined by the value of the remaining diagonal matrix elements. For the central peak one finds
\begin{equation}\label{4.5}
  \sigma G^0_{\sigma\sigma} >0\,,
\end{equation}
while for the sound peak diagonals, $G^1_{-1-1} =  -G^{-1}_{11}$, $G^1_{00} =  -G^{-1}_{00}$,   there seems to be no particular restriction. In principle, there could 
be sort of accidental zeros of $G^\alpha_{\gamma\gamma}$ which are then  difficult to locate. A more direct approach starts from the observation that 
the $\vec{G}$ coefficients are expressed through cumulants in $r_0$, $V_0$ . If the integrands are antisymmetric under reflection, many terms vanish.  The precise condition on the potential is to have some $a_0$,
$P_0$ such that
\begin{equation}\label{4.6}
V(x -a_0) +P_0x = V(-x-a_0) -P_0x\,
\end{equation}
for all $x$. Then for  arbitrary $\beta$ and $P=P_0$, one finds 
\begin{equation}\label{4.7}
G^1_{11} =0\,,\quad G^1_{-1-1} =  -G^{-1}_{11} = 0\,,\quad G^1_{00} =  -G^{-1}_{00} = 0\,,
\end{equation}
while $\sigma G^\sigma_{0\sigma'} > 0$. The standard examples for \eqref{4.6} to hold are the FPU chain with no cubic interaction term, the $\beta$-chain, and the square well potential with alternating masses, both at zero pressure. 

Under \eqref{4.7} the heat  mode is coupled to the sound mode, but there is no back reaction from the sound mode. Hence the sound peak is diffusive with  scaling function
\begin{equation}\label{4.8}
f_\sigma(x,t)=\frac{1}{\sqrt{4\pi D_\mathrm{s} t}}\mathrm{e}^{-(x-\sigma ct)^2/4 D_\mathrm{s} t}\,.
\end{equation}
$D_s$ is a transport coefficient. It can defined through a Green-Kubo formula, which also means that no reasonably explicit answer can be expected. 
The feed back of the sound peak to the central peak follows by the same computation as before, with the result   
\begin{equation}\label{4.9}
\hat{f}_0(k,t)=\mathrm{e}^{-|k|^{3/2} \lambda_\mathrm{h}t}\,,
\end{equation}
where
\begin{equation}\label{4.10}
\lambda_\mathrm{h}= (D_\sigma)^{-1/2} (G^0_{\sigma\sigma})^2 (4\pi)^2(2\pi c)^{-1/2} \int^\infty_0 dt \,t^{-1/2} \cos (t)
(2\sqrt{\pi})^{-1}\,.
\end{equation}
Since $3/2 < 5/3$, the density $f_0(x,t)$ turns out to be broader than the L\'{e}vy $5/3$ from the dynamical phase  with $G^1_{11} \neq 0$.

\begin{figure}
\includegraphics[width=1\textwidth]{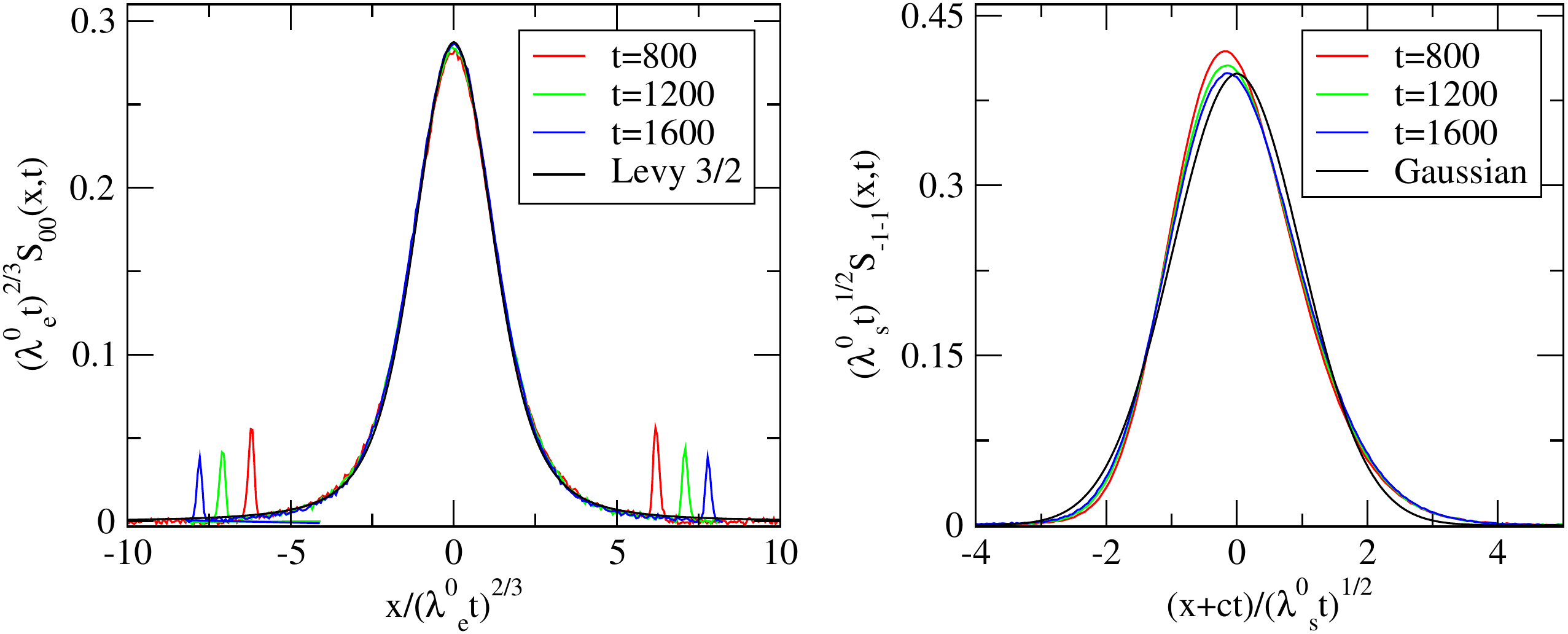}
\caption{Scaling plot of heat and sound peak for a FPU chain with $N=8192$, potential parameters $\alpha = 0$, $\beta = 1$,
pressure $P=1$, and temperature $\beta^{-1} = 1$.} 
\label{figlabel4}
\end{figure}

In testing nonlinear fluctuating hydrodynamics almost subconsciously 
 one tries to confirm (or not) the
scaling exponents, resp. functions. This can be difficult because of limited size. The dynamical phase diagram offers a different option.
For  exceptional points in the phase diagram, without too precise a verification of the scaling, one should find
that the standard scaling exponent does not  properly fit the data. Such qualitative property is 
possibly more easy to access. In Fig. \ref{figlabel4} we display heat and sound peak for a FPU chain with $G^1_{11}= 0$. 

\section{Molecular dynamics simulations}\label{sec5}
 In 1953 Fermi, Pasta and 
Ulam, technically supported by  Tsingou, simulated 32 particles interacting through a quartic potential at the 
extremely low energy of $\mathsf{e} = 5 \times 10^{-4}$ per particle (above the ground state energy) \cite{FPU56}. They injected energy in the highest Fourier mode and were looking for equipartition of the modes  at long times. However they found quasi-periodic motion with time averages settling to some definite value different from equipartition. The observed quasi-periodicity triggered the connection to KAM tori, the discovery of integrable systems with many degrees of freedom, and to the development of the theory of solitons and breathers \cite{FPU1,FPU2}.
For sure, a rich harvest, see \cite{DPR05} for a historical perspective. Later on Izrailev and Chirikov \cite{IzCh66}
repeated the simulation at the higher  energy $\mathsf{e} = 5 \times 10^{-2}$ and observed equipartition.

Anomalous transport surfaced much later \cite{LL97}, see the reviews \cite{LeLi03,Dh08}. One connects the two ends of the chain to thermal reservoirs.
To explain, the end particles are tied down as $q_0 = 0$, $q_{N+1} = 0$, and to the equations of motion for the boundary particles 
one adds Langevin terms as
\begin{equation}\label{5.1}
\ddot{ q}_1 = -\gamma p_1 + \sqrt{2\gamma T_- } \xi_-\,, \quad \ddot{ q}_N = -\gamma p_N + 
\sqrt{2\gamma T_+ } \xi_+\,,
\end{equation}
where $\gamma$ is a friction constant, $T_\pm$ are the boundary temperatures, and $\xi_\pm(t)$ are independent
standard Gaussian white noises. For $T_- = T_+$ the system settles in the canonical equilibrium state. But for 
$T_- \neq T_+$ there is a non-trivial steady state with a non-zero energy flux  $j_\mathsf{e}(N)$ depending on the length, $N$,
of the chain. For regular heat transport, Fourier's law implies $j_\mathsf{e}(N) \simeq c_0N^{-1}$. However for FPU chains one finds an enhanced transport as
\begin{equation}\label{5.2}
j_\mathsf{e}(N) \simeq c_0 N^{-1 +\alpha}(T_- - T_+)
\end{equation}
with an exponent $\alpha$ characterizing the anomaly. [A further $\alpha$, but better to stick to standard conventions]. Over the last two decades many MD simulations have been implemented for a wide variety of one-dimensional systems. Early results indicated $\alpha = 2/5$, but since about 2003 
the common evidence pointed towards $\alpha = 1/3$ or at least close to it.

Nonlinear fluctuating dynamics can also deal with such open chains, at least in principle. One would impose energy imbalance  boundary conditions as
$\ell(0,t) = 0 = \ell(L,t)$, $\mathsf{u}(0,t) = 0  =\mathsf{u}(L,t)$, but $\mathfrak{e}(0,t) =  \mathfrak{e}_-$ and 
$ \mathfrak{e}(L,t) = \mathfrak{e}_+$ and tries to investigate the steady state. Unfortunately, at least for the moment, we have no 
powerful techniques to deal with this problem. On the other hand it is argued \cite{LeLi03} that the energy flux
is related to a Green-Kubo formula by
 \begin{equation}\label{5.3}
j_\mathsf{e}(N) \sim \int _0^{N/c} dt\big(\langle \mathcal{J}_3(t);\mathcal{J}_3(0) \rangle - \langle \mathcal{J}_3(\infty);\mathcal{J}_3(0) \rangle\big)\,.
\end{equation}
Under the time integral appears the total energy current correlation in thermal equilibrium with its possibly non-zero value at $t = \infty$ subtracted, see Section \ref{sec6} for more explanations. The decay of such correlation can be predicted by mode-coupling. In fact, we will confirm the value $\alpha = 1/3$. But the argument is subtle because it is only indirectly related
to the spreading of the heat peak which is on scale $t^{3/5}$. 

Because of the relation \eqref{5.3}, in many MD simulations the total energy current correlation $\langle \mathcal{J}_3(t);\mathcal{J}_3(0) \rangle$ is measured as an addition to steady state transport
\cite{Ha99,GNY02}. There are also MD simulations exclusively focussed on momentum and energy current correlations \cite{LNG05}. However simulations of correlations 
of the conserved fields have been fairly scarce until recently. The peak structure was noted already in \cite{Pr04},
see also \cite{Zh06,Zh13}. But surprisingly enough, even 
such a basic issue as the quantitative comparison between the measured speed of sound and formula \eqref{2.27} is apparently not a routine check. So far there have been three independent sets of simulations with the specific aim to check the predictions from mode-coupling. For the details the reader is encouraged to 
look at the original papers. I will try to compare so to reach some sort of conclusion. 

 The first and second set are FPU chains with either symmetric or asymmetric potential, both the $\alpha\beta$ and the pure $\beta$ chain \cite{St13,DDSMS14}. In this case one has to integrate numerically the differential equations governing the evolution, for which both a velocity-Verlet algorithm and a fourth order symplectic Runge-Kutta algorithm are used. The third set consists of chains with a piecewise constant potential \cite{MeSp14}. We call them hard-collision, since the force is zero
except for $\delta$-spikes and the dynamics proceeds from collision to collision. Now one has to develop an efficient algorithm by which one finds the time-wise next collision. Except for rounding, there is no discretization of time. 
The simplest example is the hard-point potential, $V_\mathrm{hc}(x) =\infty$ for $x<0$ and $V_\mathrm{hc}(x) = 0$ for $x\geq 0$. A variant is the infinite square well potential, $V_\mathrm{sw}(x) =\infty$ for $x<0$, $x >a$ and $V_\mathrm{sw}(x) = 0$ for $0 \leq x\leq a$ \cite{LiLe05,MeSp14}. In this case two neighboring particles at the maximal distance $a$ are reflected inwards as if connected by a massless string of finite length $a$. For both models the 
dynamics remembers the initial velocities. To have only the standard conservation laws, one imposes  alternating masses, say $m_j = m_0$ for even $j$ and $m_j = m_1$ for odd $j$. In both models  the unit cell then contains two particles and the scheme explained before has to be extended. But at the very end the difference is minimal. A further variant is the shoulder
potential $V_\mathrm{sh}(x) $, for which $V_\mathrm{sh}(x) = \infty $ for $|x| < \tfrac{1}{2}$, $ V_\mathrm{sh}(x)  = \epsilon_0$ for
$\tfrac{1}{2} \leq |x| \leq 1$, and  $V_\mathrm{sh}(x) = 0 $ for $|x| > 1$. The potential is either repulsive, $\epsilon_0 > 0$, or attractive $\epsilon_0 < 0$. Exploratory studies of the latter case indicate that the convergence is slower than for the extensivlely studied attractive case. Particles interacting with such a potential can be viewed
also as a hard-core fluid with a short range potential part and thus serves as a bridge between one-dimensional fluids and anharmonic solids. The collisions resulting from the potential step make the model non-integrable.

FPU chains with an even potential at $P= 0$ constitute a distinct dynamical phase. Such phase is absent for the hard-point and the square shoulder potentials. But the square well at $P=0$ has the same properties as can be seen from taking $a_0 = a/2$ in \eqref{4.6}.

Current system sizes are $N = 2^{11}$ to $2^{13}$, even size $2^{16} = 65,536$ has been attempted \cite{DDO13}. Periodic boundary conditions are imposed. Time is restricted to $ t \leq t_\mathrm{max} =  N/2c$, the time of 
the first collision between sound peaks. For given potential, one has to decide on the thermodynamic parameters,
$P,\beta$. One constraint is to have $c$ approximately in the range $1\,...\,2$ in order to have a sufficiently long time span available.
Secondly one would like to be well away from integrability. This leads to an energy per particle of order 1 in the models from above. A related issue are the coupling matrices $\vec{G}$, which should not be too small, at least for the relevant couplings. For the models from above they are tabulated. The relevant couplings show quite some variability taking values in the range $0.1\, ... \,3.4$. At the very end, one has to make a physically reasonable choice, perhaps use the same parameters as previous MD simulations so to have the possibility to compare. A systematic study of the dependence on $P,\beta$ seems to be too costly and most likely not so interesting. But it does make sense to probe values at the border. For some of the FPU simulations the energy per particle is chosen
as
$\mathsf{e} = 0.1$, at which point nonlinearities are small \cite{St13}. There are also very extensive simulations at the even lower energy
$\mathsf{e} = 5 \times 10^{-4}$, with the goal to explore  the route to equipartition, which is a somewhat distinct issue
\cite{Be12}.

Once all parameters are fixed, there are several options to run the simulation. Since in canonical equilibrium 
$\{r_j,p_j, j = 1,...,N\}$ are independent random variables, one can sample the initial conditions through a random number generator.
For the hard-collision potentials the geometric constraints are still simple enough for allowing one to generate the microcanonical ensemble by Monte Carlo methods. In our simulations the correlator $S$ hardly depends on the choice of the ensemble. With such generated random initial data the equations of motion are simulated up to  $t_\mathrm{max}$. A single run is noisy and one has to repeat many times, order $10^7$.
The much more common choice is to start from a reasonable nonequilibrium configuration and to equilibrate
before measuring correlations. Usually one then simulates very long trajectories, up to times of order $2^{15}$, over which the time lag $g_\alpha(j,t+\tau)g_{\alpha'}(0,t)$ is sampled, $\tau \leq t_\mathrm{max}$. In addition one  averages over a small number of runs, of order $10^2$. The total number of samples is roughly the same in both methods. The random number generator method produces  the thermal average with a higher reliability.

The sampled $S_{\alpha\alpha'}(j,t)$ can be Fourier transformed in the spatial variable and/or in the time variable.
One can also transform to normal coordinates. These are linear operations which can be done for each sample or only after averaging. Should one keep the full resolution or only some data points? Of course it depends somewhat on the goals. In \cite{DDSMS14,MeSp14}, the full $3 \times 3$ matrix is sampled and subsequently transformed via the theoretically computed  $R$ matrix to obtain
$S^\sharp(j,t)$. This approach allows to test diagonality. Because the peak structure is most easily seen in the space-like $j$ coordinate,
maximal resolution for $j$ is retained and only three times $(t = 250,500, 1000)$ are recorded for the purpose of making a scaling plot. 
In \cite{vBPo12}, the lowest Fourier modes are measured as a function of $t$. In \cite{St13} the lowest Fourier modes are plotted 
as a function of the frequency $\omega$. A separate issue are the much simulated total current correlations. The
total currents are sampled directly, in the most complete version momentum, energy, and cross correlations, and then plotted as a function of $t$ or $\omega$. 
\begin{figure}
\includegraphics[width=0.49\columnwidth]{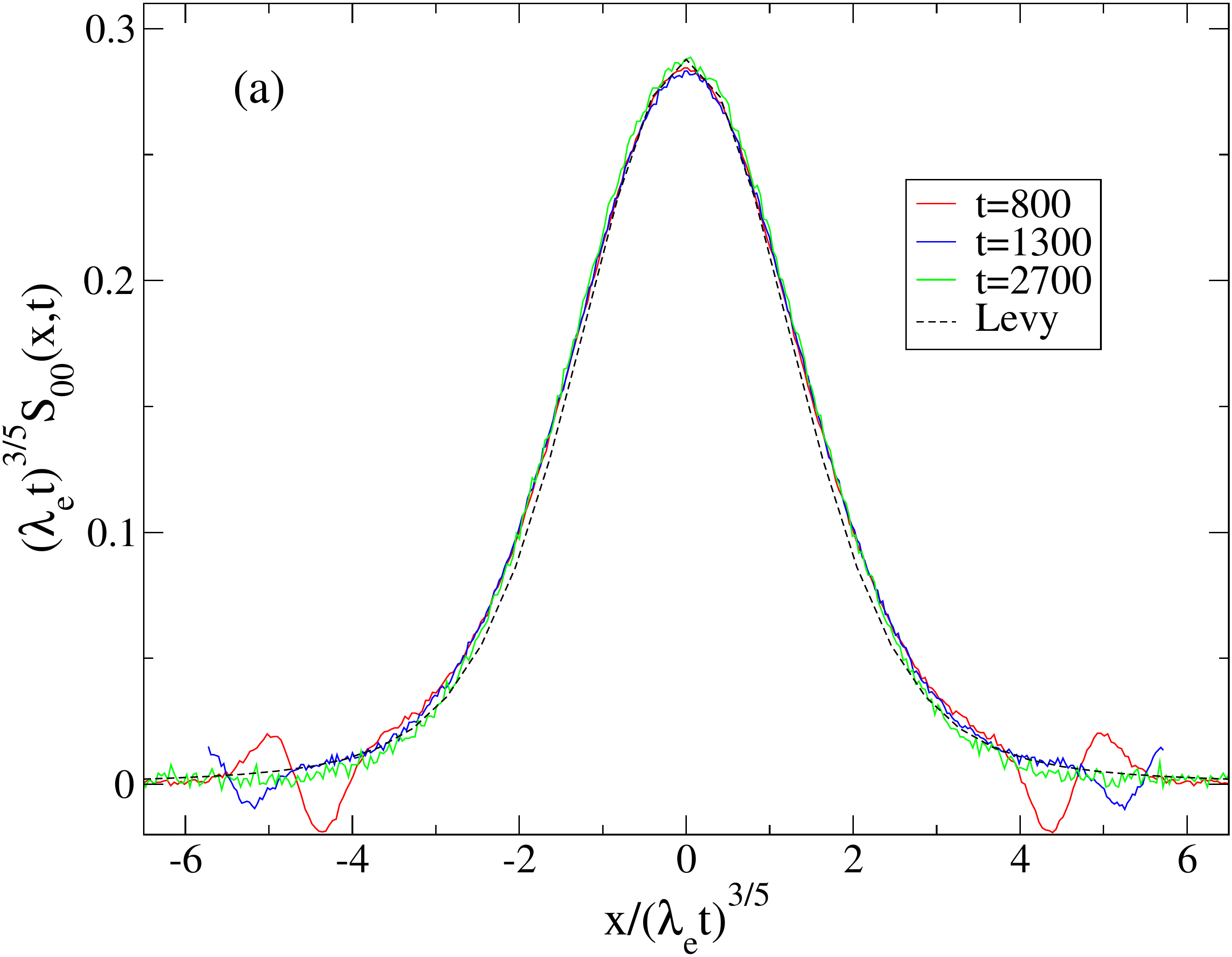} 
\includegraphics[width=0.49\columnwidth]{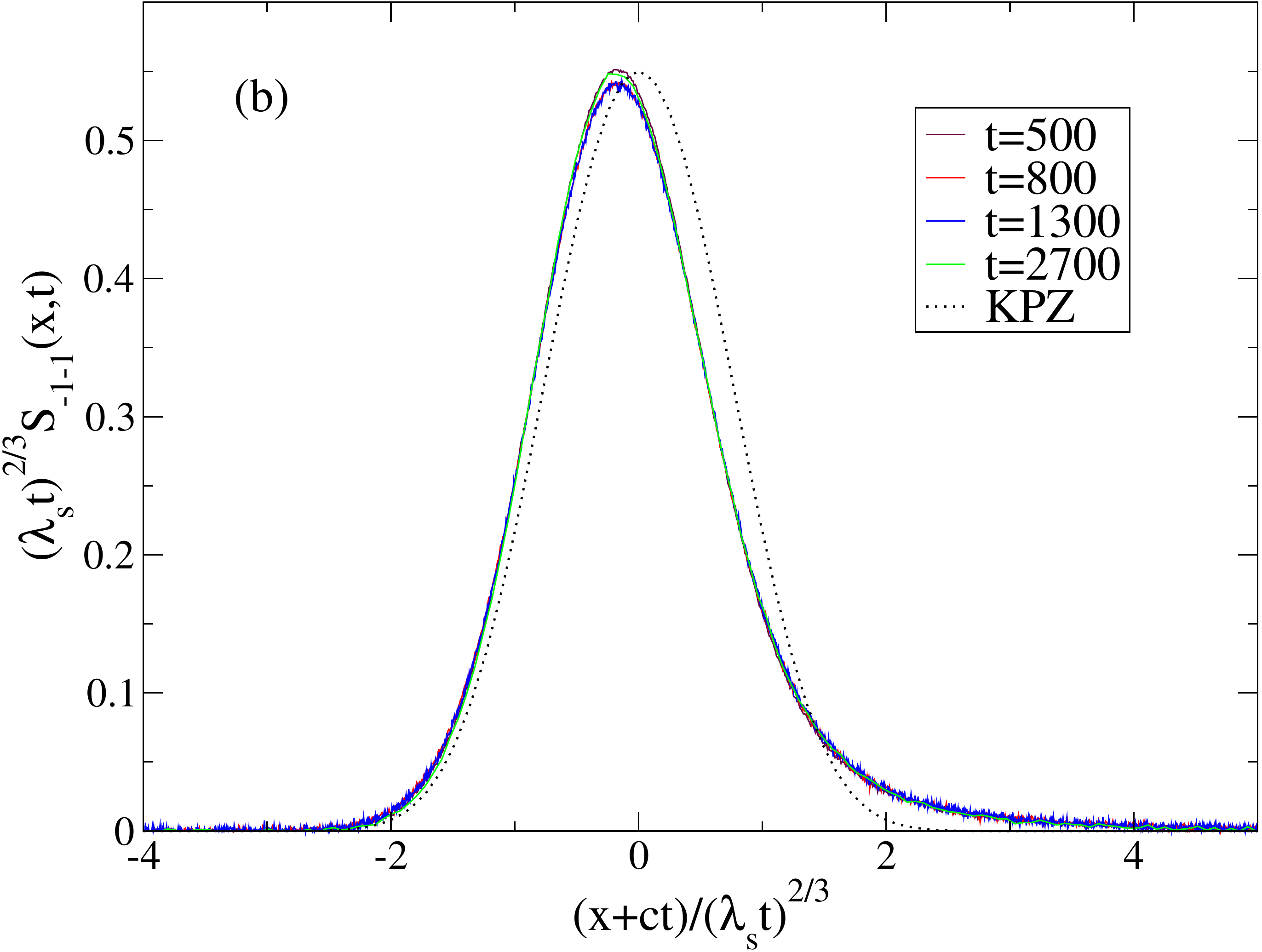}
\caption{ Scaling plot of heat and sound peak  for a FPU chain with $N=8192$, potential parameters $\alpha = 2$, $\beta = 1$,
pressure $P=1$, and temperature $\beta^{-1} = 0.5$.}
\label{figlabel5}
\end{figure}

We reproduce only a few figures. Many more details can be found in the original papers. In Fig. \ref{figlabel5} we display the data for the FPU chain 
\cite{DDSMS14} with
$V_\mathrm{FPU}(x) = \tfrac{1}{2}x^2+ \tfrac{2}{3}x_3 + \tfrac{1}{4}x^4$, $\beta = 2$, $P=1$, $N = 8192$, and $t_\mathrm{max} 
= 2700$.  The sound speed is $c = 1.45$. Note that the sound peak is somewhat distorted, not symmetric 
relative to $ct$, but has a rapid fall-off away from the sound cone. As only fit parameter one uses $\lambda_{\mathrm{h}}$, resp. $\lambda_\mathrm{s}$. The optimal fit at the longest available time is denoted by  $\lambda^\mathrm{emp}_{\mathrm{h}}$, resp. $\lambda^\mathrm{emp}_\mathrm{s}$, standing for empirical value. In most cases there is also a theoretical value based on decoupling and/or mode-coupling, which is indicated  in square brackets. For the FPU simulations the results are
for the heat peak $\lambda^\mathrm{emp}_{\mathrm{h}} = 13.8\, \,[1.97]$,
and for the sound peak   $\lambda^\mathrm{emp}_\mathrm{s} = 2.05\, \,[0.68]$. From the visual appearance, one might have guessed the theoretical values to be just the other way round. So maybe the system tries to generate the optimal L\'{e}vy  peak at non-relaxed sound peaks.
 \begin{figure}[!htp]
\centering
\subfloat[shoulder, heat]{
\includegraphics[width=0.33\columnwidth]{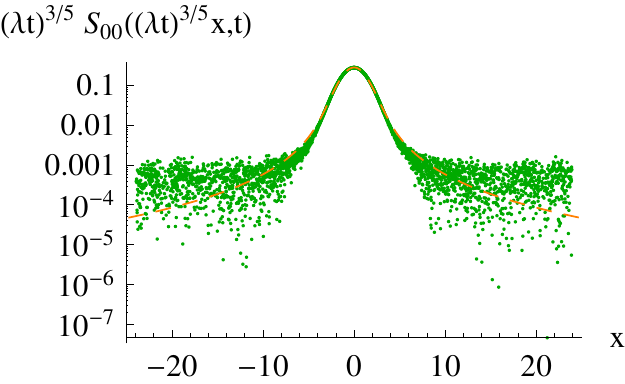}}
\subfloat[hard-point gas, heat]{
\includegraphics[width=0.33\columnwidth]{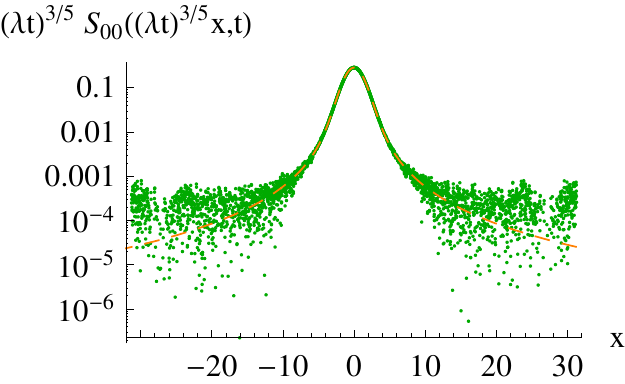}}
\subfloat[square-well, heat]{
\includegraphics[width=0.33\columnwidth]{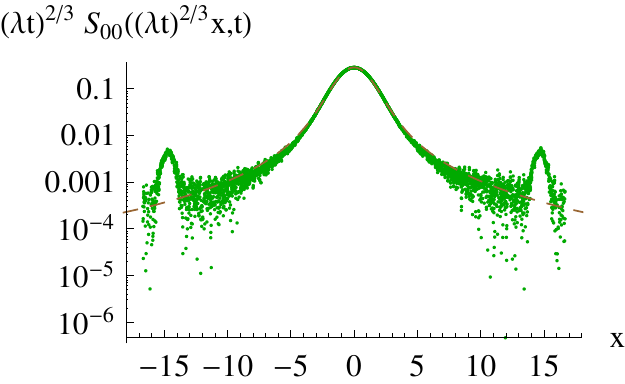}} \\
\subfloat[shoulder, sound]{
\includegraphics[width=0.33\columnwidth]{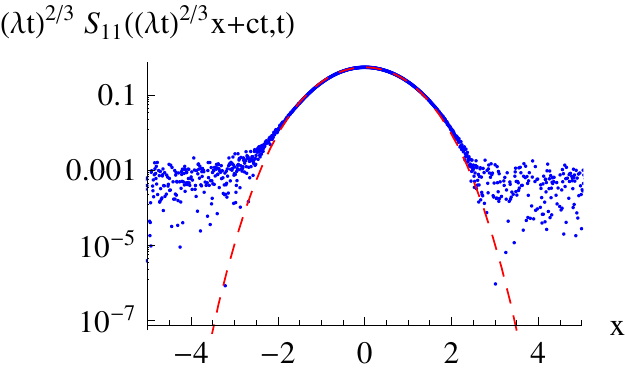}}
\subfloat[hard-point gas, sound]{
\includegraphics[width=0.33\columnwidth]{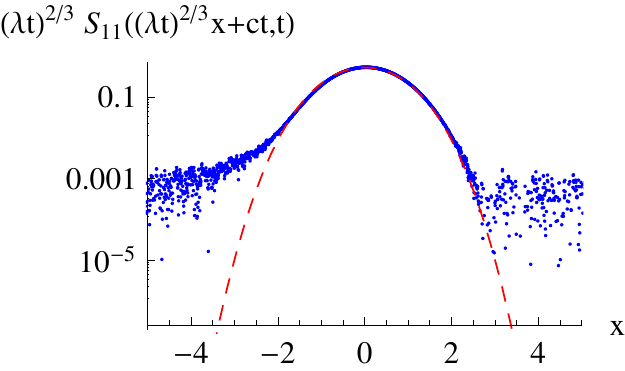}}
\subfloat[square-well, sound]{
\includegraphics[width=0.33\columnwidth]{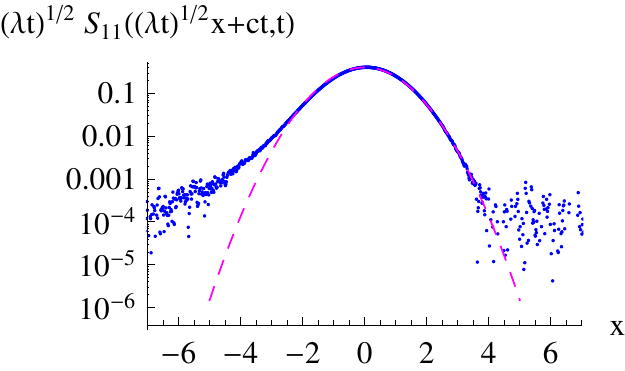}}
\caption{Heat and sound peak for shoulder, hard-point, and square well potential in logarithmic scale.}
\label{figlabel1}
\end{figure}

In Fig. \ref{figlabel1} we reproduce the plot of  heat and sound peak for the hard-collision models with shoulder, hard-point, and square well potential, in the latter two cases with alternating masses \cite{MeSp14}.  To better judge the quality of the fit we provide a logarithmic plot. In all cases $N=4096$ with $t_\mathrm{max} = 1024$. The following parameters have been chosen, shoulder: $\epsilon_0 = 1$, $P = 1.2$, $\beta = 2$, $c= 1.74$, hard-point: $m_1/m_0 = 3$, $P = 2$, $\beta = 0.5$, $c= 1.73$, square well:
$m_1/m_0 = 3$, $a=1$, $P = 0$, $\beta = 2$, $c= 1.73$. The fit to the predicted scaling function has an error less than $5\%$. For the hard-collision models the simulation results are,  shoulder: $\lambda^\mathrm{emp}_\mathrm{s} = 1.62 \,\,[1.71]$, $\lambda^\mathrm{emp}_\mathrm{h} = 1.44\,\, [1.04]$, hard-point: $\lambda^\mathrm{emp}_\mathrm{s} = 1.42 \,\,[2.00]$, $\lambda^\mathrm{emp}_\mathrm{h} = 1.04\, \,[0.95]$, square well: $\lambda^\mathrm{emp}_\mathrm{h} = 0.95\, \,[1.04]$. Recall that the square well potential at $P=0$ is in a distinct dynamical universality class, hence the different scaling
exponents. For this model   $\lambda_\mathrm{s}$ is related to a diffusion constant,
which can be obtained only numerically.
  
\begin{figure}[!ht]
\centering
\includegraphics[width=0.7\textwidth]{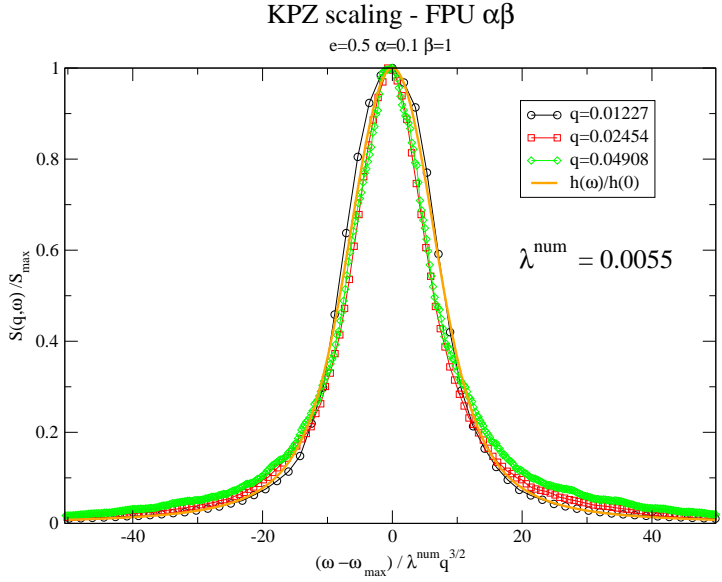}
\caption{Scaling plot of  the sound peak in $(k = q,\omega)$ variables of a FPU chain with $N = 1024$, potential parameters
$\alpha = 0.1$, $\beta = 1$, pressure $P = -0.04$, and temperature $\beta^{-1} = 0.105$. The speed of sound is $c= 1.11$.}
\label{figlabel6}
\end{figure}

The MD simulation \cite{St13} is at low temperatures and considers the positional correlations, which in Fourier space differ from the stretch correlations only by a $k$-dependent prefactor. Thus sampled is the correlator
\begin{equation}\label{5.4}  (1- \cos(2\pi k))^{-1}\int dt\,\mathrm{e}^{\mathrm{i}\omega t}\hat{S}_{11}(k,t) = (1- \cos(2\pi k))^{-1}\hat{S}_{11}(k,\omega)\,.
\end{equation}
For low temperatures the area under the central peak is a factor $10$ smaller than the one under the sound peak. Hence only the sound peak is explored. Its
asymptotic scaling form is 
\begin{equation}\label{5.5}
\hat{S}_{11}(k,t) = \cos(2\pi \mathrm{i}kct)  
C_{11}\hat{f}_{\mathrm{KPZ}}(k(\lambda_\mathrm{h}|t|)^{2/3})\,.
\end{equation}
Considering only the right moving sound peak by setting $\omega_{\mathrm{max}} = 2\pi kc$,
\begin{eqnarray}\label{5.6}
&&\hspace{-20pt}\hat{S}_{11}(k,\omega+\omega_{\mathrm{max}}) = \int dt\,\mathrm{e}^{\mathrm{i}\omega t} \tfrac{1}{2} C_{11} \hat{f}_{\mathrm{KPZ}}
(k(\lambda_\mathrm{h} |t|)^{2/3})\nonumber\\
&&\hspace{20pt}= \int dt\,\mathrm{e}^{\mathrm{i}(\omega/\lambda_\mathrm{h} |k|^{3/2}) t} (\lambda_\mathrm{h}  |k|^{3/2})^{-1}
\tfrac{1}{2} C_{11}\hat{f}_{\mathrm{KPZ}}(|t|^{2/3})\,.
\end{eqnarray}
Thus defining
 \begin{equation}\label{5.7}
h_{\mathrm{KPZ}}(\omega) = \int dt\,\mathrm{e}^{\mathrm{i}\omega t}  \hat{f}_{\mathrm{KPZ}}(|t|^{2/3})\,,
\end{equation}
one arrives at
 \begin{equation}\label{5.8}
\hat{S}_{11}(k,\omega+\omega_{\mathrm{max}}) = \tfrac{1}{2} C_{11} (\lambda |k|^{3/2})^{-1} h_{\mathrm{KPZ}}(\omega/ \lambda_{11} |k|^{3/2})\,.
\end{equation}
If one normalizes the maximum of $\hat{S}$ to 1, then $h_{\mathrm{KPZ}}(\omega)$ is replaced by $h_{\mathrm{KPZ}}(\omega)/h_{\mathrm{KPZ}}(0)$, which amounts to setting the prefactor in (\ref{5.8}) equal to 1. In Fig. \ref{figlabel6}
the spectrum at two different choices of the asymmetry parameter is displayed.

In the simulation the potential is chosen as $V(x) = \tfrac{1}{2}x^2 + \alpha \tfrac{1}{3}x^3 + \tfrac{1}{4}x^4$, where the asymmetry  varies from $0$ to $2$. Increasing $\alpha$, the pressure $P$ increases from $0$ to $0.2$ and the inverse temperature   from $9.55$ to $9.75$. The sound speed $c \simeq 1.1$. In frequency space the peak moves linearly in $k$ at around $\omega = 0.01$. For the shape function one uses $\hat{S}(k,\omega)$ for $k = 1,2,4,8,16$ to generate  a scaling plot.
Over the whole range of $\alpha$'s the fit with the scaling function \eqref{5.7} is fairly convincing. The optimal fit  parameter starts  from $\lambda^\mathrm{emp}_\mathrm{s} = 0.02 \,\,[0.04]$ at $\alpha= 0.2 $ to  $\lambda^\mathrm{emp}_\mathrm{s} = 0.08 \,\,[0.37]$ 
at $\alpha= 1.6$ and to
$\lambda^\mathrm{emp}_\mathrm{s} = 0.07 \,\,[0.53]$ at $\alpha= 2.0$.

Given the diversity of models, parameters,  and numerical schemes only tentative conclusions can be drawn. \medskip\\
\textit{(i)} The separation into three peaks is a fast process. In normal mode representation the off-diagonal matrix elements are indeed small. There are no correlations beyond the sound cone.\\
\textit{(ii)} The distinction between $G^1_{11} \neq 0$ and  $G^1_{11} = 0$  is seen very convincingly.\\
\textit{(iii)} The central peak is on the $t^{3/5}$ scale and adjusts well to the predicted L\'{e}vy distribution with one caveat.
As seen from the center, the shape function fairly rapidly switches into the power law decay. However at the location of the sound peaks there are some wiggles and beyond only small amplitude noise is observed. In this sense, the L\'{e}vy distribution gets uncovered as time progresses. The L\'{e}vy distribution is a fairly direct consequence of  $G^0_{00} = 0$
and thus one of the strongest supports for the theory. \\
\textit{(iv)} The sound peaks are mirror images of each other. One plots them on the $t^{2/3}$ scale, but then there is still
a slow change in time. For the hard-collision models the shape is almost perfect, but the  $\lambda_\mathrm{s}$
parameter is dropping in time. For the FPU models the peak is distorted and still away from the symmetric shape predicted by the theory. To the outside of the sound cone there is the rapid fall-off in accordance with the KPZ scaling function. But towards the heat peak there is slow decay. The sound peak is tilted away from the central peak. Apparently there is still a strong interaction between the peaks.\\
\textit{(v)} In the majority of the simulations the peaks vary slowly on the scale $t^{2/3}$ for sound, resp. $t^{3/5}$ for the heat peak.
Thus it becomes meaningful to use as a fit the theoretical scaling function with $\lambda_{\mathrm{s}}$, resp. $\lambda_\mathrm{h}$,
as only free parameter. This optimal choice has been denoted above as empirical value $\lambda^\mathrm{emp}_{\mathrm{s/h}}$. The error between the measured and theoretical shape function is  less than $5\%$.
However one observes that $\lambda^\mathrm{emp}_{\mathrm{s}}$ and $\lambda^\mathrm{emp}_{\mathrm{h}}$ are still changing in time signaling that the simulation has not yet reached the truly asymptotic regime. In some simulations   $\lambda^\mathrm{emp}_{\mathrm{s/h}}$  drops monotonically in time and differs not too strongly from 
$\lambda_{\mathrm{s/h}}$. One is then willing to believe that for even longer simulation times the
asymptotic value is reached. But in other simulations there is a much stronger discrepancy, which  asks for more explanations.

Mode-coupling is not specific to anharmonic chains. In principle any one-dimensional system with conserved fields can be 
handled by the same scheme. This offers the possibility to test the theory through other models, possibly finding systems with less strong finite time effects. One obvious choice are stochastic lattice gases with several type of particles
like several lane TASEP \cite{PSS14,PSSS15} and the AHR model \cite{AHR99,FeSS13}. For them the couplings can be more easily adjusted than for anharmonic chains, which offers the possibility to test the dynamical phase diagram.
Also anharmonic chains with a stochastic collision mechanism, respecting the conservation laws, have been studied in considerable  detail \cite{BGJ14,JKO14,StSp14}.  
\section{Total current correlations}\label{sec6}
The total current is a fluctuation observable, in contrast to $S_{\alpha\alpha'}(j,t)$ which refers to the average of the product of two local observables. Thus we need some additional considerations to establish the link to nonlinear fluctuating hydrodynamics.
For a ring of size $N$, $\Lambda_N = [1,...,N]$, the total currents are defined by 
\begin{equation}\label{6.1}
\vec{\mathcal{J}}_{\mathrm{tot},\Lambda_N}(t)   = \frac{1}{\sqrt{N}} \sum_{j=1}^{N} \vec{\mathcal{J}}(j,t )   
\end{equation} 
and the total current covariance reads
\begin{equation}\label{6.2}
\Gamma_{\Lambda_N,\alpha\alpha'} (t) =   \langle \mathcal{J}_{\mathrm{tot},\Lambda_N,\alpha}(t) ; \mathcal{J}_{\mathrm{tot},\Lambda_N,{\alpha'}}(0) \rangle_{P,\beta,\Lambda_N}
 =  \sum_{j=1}^{N} \langle \mathcal{J}_\alpha(j,t );\mathcal{J}_{\alpha'}(0,0 )\rangle_{P,\beta,\Lambda_N}\,.
\end{equation} 
The cumulant $\langle\cdot;\cdot\rangle$ means that the static average is subtracted and system size is indicated explicitly. In the limit $N \to \infty$
\begin{equation}\label{6.2a}
\Gamma_{\alpha\alpha'} (t)
 =  \sum_{j\in\mathbb{Z}} \langle \mathcal{J}_\alpha(j,t );\mathcal{J}_{\alpha'}(0,0 )\rangle_{P,\beta}\,.
\end{equation} 
For fixed $t$ the integrand decays exponentially in $j$, but with a correlation length increasing in time.

\begin{figure}[!htp]
\centering
\subfloat[momentum current correlations]{
\includegraphics[width=0.4\textwidth]{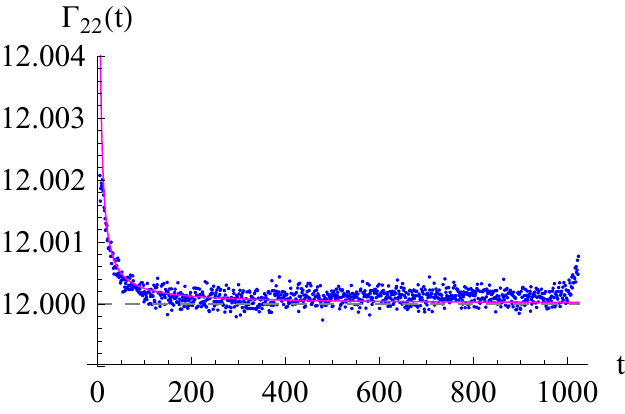}}
\hspace{0.01\textwidth}
\subfloat[logarithmic plot of $\Gamma^{\scriptscriptstyle\Delta}_{22}(t)$]{
\includegraphics[width=0.4\textwidth]{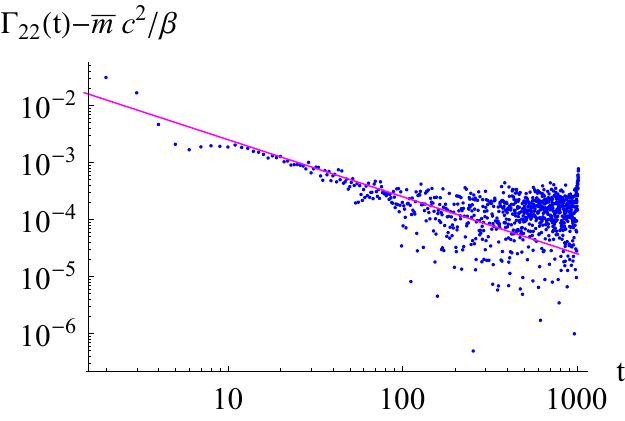}}\\
\subfloat[energy current correlations]{
\includegraphics[width=0.4\textwidth]{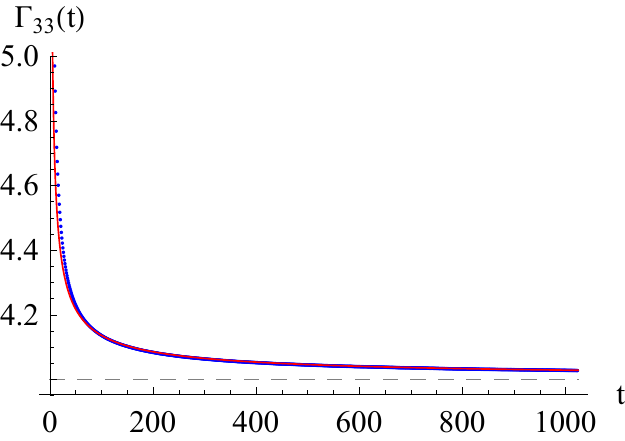}}
\hspace{0.01\textwidth}
\subfloat[logarithmic plot of $\Gamma^{\scriptscriptstyle\Delta}_{33}(t)$]{
\includegraphics[width=0.4\textwidth]{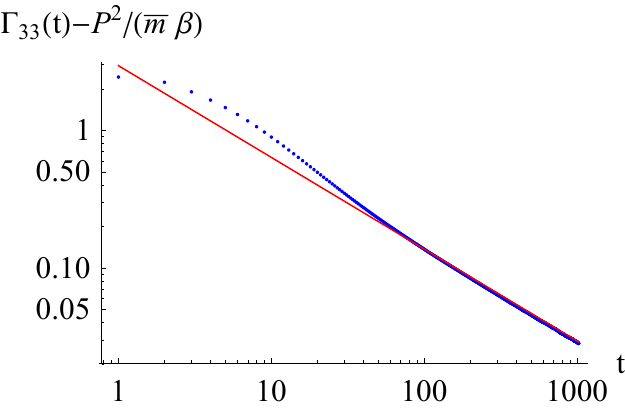}}
\caption{Total momentum and energy correlations for a hard-point particles with alternating masses}
\label{figlabel2}
\end{figure}

Before we emphasized the transformation to normal coordinates. For the currents we stick to the physical fields. The current $\mathcal{J}_{\mathrm{tot},\Lambda_N,1}(t)$ is itself conserved. Thus only the $(2,3)$ block has a time variation. 
Using stationarity and time-reversal, the diagonal elements are even, 
$\Gamma_{\alpha\alpha}(t) = \Gamma_{\alpha\alpha}(-t)$, while the off-diagonal elements are odd and satisfy 
 \begin{equation}\label{6.3}
\Gamma_{23} (t) = -\Gamma_{23} (-t) = \Gamma_{32} (- t)= -\Gamma_{32}(t)\,.
\end{equation} 
Mode-coupling predicts that this matrix element vanishes. In our simulations we observe an exponential decay
with a decay time of order 20 to 30. Thus the correlations of real interest are $\Gamma_{22} (t)$
and $\Gamma_{33} (t)$. For chains the latter is the most frequently simulated equilibrium time correlation. The momentum current correlations have been measured  in \cite{LNG05,vBPo14a} and the momentum-energy cross correlations only recently in \cite{MeSp15}.

$\Gamma_{\Lambda_N,\alpha\alpha'} (t)$ is a fluctuation
observable, for which the equivalence of ensembles does not hold. But the time-independent difference between microcanonical and canonical average can be computed explicitly. For the microcanonical ensemble, $\lim_{t \to \infty} \Gamma_{\Lambda_N,\mathrm{micro}}(t) = 0$. On the other hand, there is no reason for $\Gamma_{\Lambda_N}(t)$ to vanish asymptotically if the canonical average,
as in \eqref{6.2}, is used. In fact, for infinite volume, 
 \begin{equation}\label{6.4}
\lim_{t \to \infty}\Gamma_{22} (t) =  \beta^{-1}c^2 \,,\qquad  
\lim_{t \to \infty}\Gamma_{33} (t)=  \beta^{-1} P^2 \,.
\end{equation} 

The asymptotic values in \eqref{6.4} are called Drude weight, which has received a lot of attention in the 
context of current correlations for integrable quantum chains. A non-zero Drude weight indicates that the
correlator of the corresponding conserved field has a ballistically moving component, but it cannot resolve
the structure of this component. For non-integrable anharmonic chains, the ballistic pieces are just the sharply concentrated sound peaks, while in the integrable case one expects to have a broad spectrum which expands ballistically.

The link to mode-coupling is achieved through the general observation that the current correlations are proportional
to the memory kernel. We use its diagonal approximation and insert the asymptotic  form of $f_\alpha$. The
memory kernel is in normal mode representation. Thus we still have to transform back to the physical fields through the $R$ matrix. The 
computation can be found in \cite{MeSp15} with the result
 \begin{eqnarray}\label{6.4a}
&&\hspace{-20pt}\Gamma^{\scriptscriptstyle\Delta}_{22} (t)  = 
\Gamma_{22} (t) - \beta^{-1}c^2 \nonumber\\
&&\hspace{0pt} \simeq \tfrac{1}{2}(\lambda_{\mathrm{h}}t)^{-3/5} 
\langle \psi_0,H^{\mathsf{u}} \psi_0\rangle^2 \int dx\, f_{\text{L\'evy},5/3}(x)^2
 + (\lambda_{\mathrm{s}}t)^{-2/3} \langle \psi_1,H^{\mathsf{u}} \psi_1\rangle^2 \int dx\, f_{\mathrm{KPZ}}(x)^2  
\nonumber\\
&&\hspace{-20pt}\Gamma^{\scriptscriptstyle\Delta}_{33} (t)= \Gamma_{33} (t)- \beta^{-1} P^2 \simeq c^2\beta^{-2}(\lambda_{\mathrm{s}}t)^{-2/3} \int dx\, f_{\mathrm{KPZ}}(x)^2\,,
\end{eqnarray} 
where $\{\psi_\alpha\}$ are the eigenvectors of $A$, $A\psi_0 = 0$, $A\psi_1 = c \psi_1$, see \cite{Sp14}. If \eqref{4.6} is satisfied, \textit{e.g.} an even potential at $P=0$, then the sound peak is diffusive, see Eq. \eqref{4.8}, and the central peak is L\'{e}vy $3/2$.  Furthermore $\langle \psi_0,H^{\mathsf{u}} \psi_0\rangle = 0 = \langle \psi_1,H^{\mathsf{u}} \psi_1\rangle$.
Hence $\Gamma_{22} (t) - \beta^{-1}c^2$ is expected to decay integrably, while 
\begin{equation}\label{3.6a}
\Gamma_{33} (t)- \beta^{-1} P^2 \simeq c^2\beta^{-2} (8\pi D_\mathrm{s} t)^{-1/2}\,.
\end{equation}

The energy current correlation is predicted to decay as $t^{-2/3}$ which has been reported in  MD simulations already 
more than 15 years ago \cite{Ha99,GNY02}. The true mechanism behind the decay is actually somewhat subtle. From the conservation law it follows that the second moment of the heat peak is related to the second time derivate of the current correlation. Using  the asymptotic form \eqref{50}, including the cut-off at the sound peak,  
one arrives at 
\begin{equation}\label{3.6b}
\frac{d^2}{dt^2} \int_{-c t}^{c t} dx\, x^2 f_0(x,t) \simeq \tfrac{8}{3\pi}( \lambda_\mathrm{h})^{5/3} c^{1/3} (\lambda_\mathrm{h} t)^{-2/3}\,.
\end{equation}
This argument overlooks that the scaling form is for the normal mode representation, while $\Gamma_{33}(t)$ refers to the physical energy current. The complete computation leads however to the same power law except for a different prefactor \cite{MeSp15}.

\begin{figure}[!htp]
\centering
\includegraphics[width=0.9\textwidth]{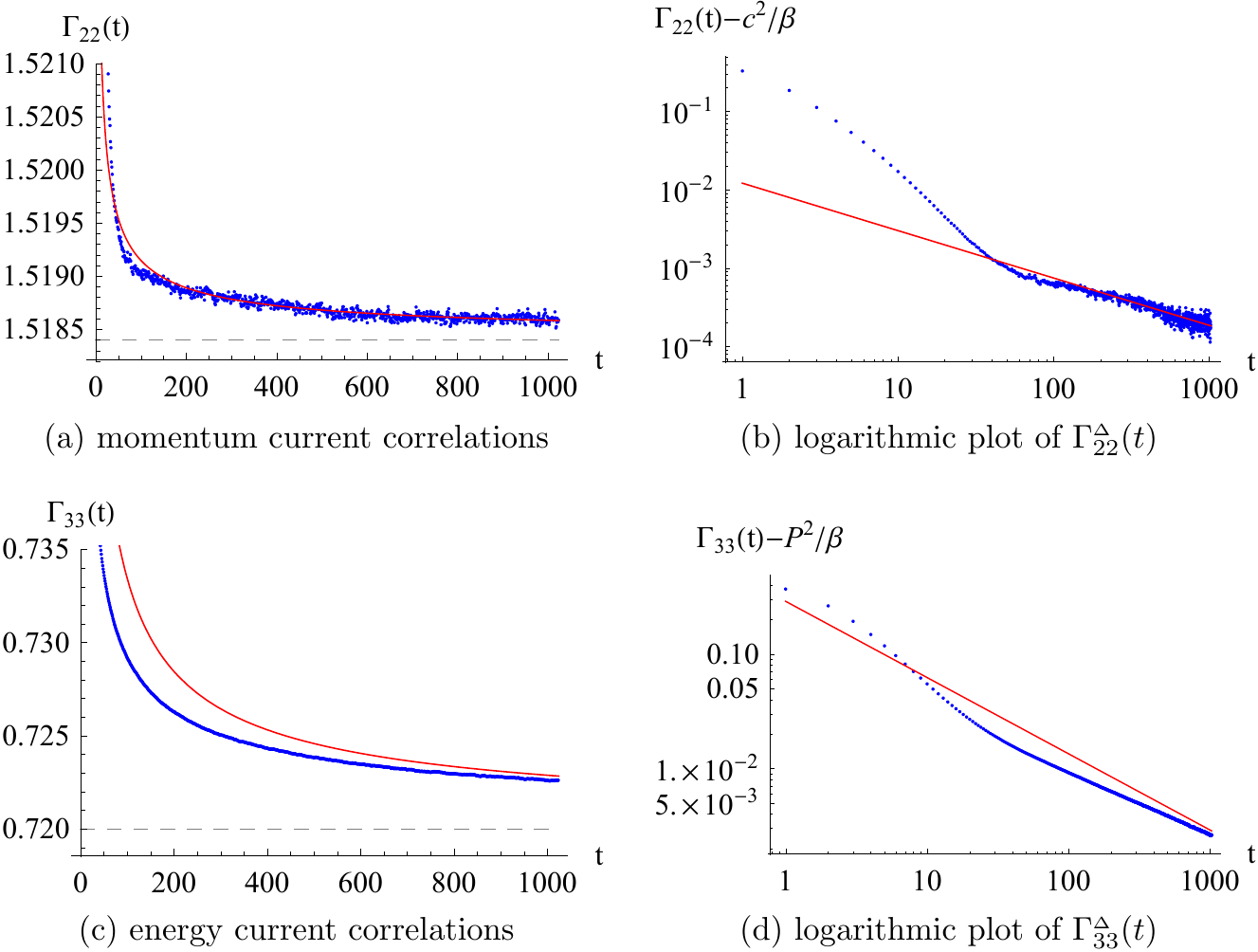}
\hspace{0.01\textwidth}
\caption{Total momentum and energy current correlations for a hard collision model with shoulder potential.}
\label{figlabel3}
\end{figure}

The momentum current correlation should decay as $t^{-3/5}$, which is a recent finding. However, its prefactor could vanish, in principle. Mode-coupling with the currently available precision would not provide an answer, then. In contrast the prefactor for $\Gamma_{33}(t)$ is strictly positive.

Our simulation results \cite{MeSp15} are  shown in Fig. \ref{figlabel2},\ref{figlabel3}. The parameters are as before, $N=4096$, $t_\mathrm{max} = 1024$,   
shoulder: $\epsilon_0 = 1$, $P = 1.2$, $\beta = 2$, $c= 1.74$, and hard-point: $m_1/m_0 = 3$, $P = 2$, $\beta = 0.5$, $c= 1.73$.
The red lines indicate the predictions based on mode-coupling.
It is interesting to note that for the  shoulder potential the evidence for a $t^{-2/3}$ decay is not so overwhelming as one might have anticipated and, by looking at a different time window, one could as well fit to a slightly different exponent.
On the other hand the hard-point potential with alternating masses shows a very clean power law decay. Through  MD with shoulder potential the predicted decay of $\Gamma_{22}(t)$ is well confirmed. However, for the hard-point potential it so happens that both prefactors, $\langle \psi_0,H^{\mathsf{u}} \psi_0\rangle$ and
 $\langle \psi_1,H^{\mathsf{u}} \psi_1\rangle$, 
  vanish. Numerically we estimate a decay as $t^{-1}$. 
  Again this is a strong qualitative support of nonlinear fluctuating hydrodynamics. One might have thought that all hard-collision potentials have the same asymptotic power law for the momentum current correlation. This expectation is born out under the proviso that the respective prefactors do not vanish.   
Matrix elements, as $\langle \psi_0,H^{\mathsf{u}} \psi_0\rangle$,  must come from a microscopic computation and 
cannot be deduced by a mere inspection of the potential.

An additional confirmation of \eqref{6.4}, \eqref{3.6a} has been accomplished recently \cite{LD15}. For the potential $V(x)
= \tfrac{1}{3}ax^3 + \tfrac{1}{4}x^4$, $ \beta = 1$ and at $P= 0.59$ with $a= -2$, resp. at $P = -0.5$, $a= 1.89$, up to very small errors the signature of the $\vec{G}$ matrices is identical to an even potential at $P=0$. In the MD simulation the energy current is
found to decay as $t^{-1/2}$ and the momentum current seems to be integrable. Keeping all parameters fixed and shifting slightly to $a = -2.7$, resp. $a=2$, the decay as stated in  \eqref{6.4a} is restored.   
\section{Other 1D Hamiltonian systems}\label{sec7}
For the anharmonic chains studied so far, the potential depends only on $q_{j+1} - q_j$ and hence 
remains without change under spatial translations. Physically this property is obvious and seems hard to avoid.
However there could be a substrate potential which forces the particles preferentially to particular locations. One could consider a two-component system,
which then has an acoustic and an optical mode. The latter would be comparable to a one-component system with an on-site potential. Such considerations lead to the more general class of Hamiltonians
\begin{equation}\label{7.1a}
H_\mathrm{os} = \sum_{j=1}^N \big(\tfrac{1}{2} p_j^2 + V_\mathrm{os}(q_j)\big) + \sum_{j = 1}^{N-1}V(q_{j+1} - q_j)
 \end{equation}
with some confining on-site potential $V_\mathrm{os}$. The only conserved field is the energy. There is a unique
equilibrium measure. The Euler currents vanish. From the perspective of fluctuating hydrodynamics all evidence points towards diffusive energy transport. For the case of a quadratic $V$ and for $V_\mathrm{os} = V_\mathrm{FPU}$
very detailed MD simulation confirm diffusive transport \cite{Aoki}.

More interesting are models with two conservation laws. We discuss separately coupled rotators, which can be thought of as a classical limit of a quantum Heisenberg chain, and the discrete nonlinear Schr\"{o}dinger equation on a lattice,
which is the classical field theory for lattice bosons.\medskip\\
\textbf{Coupled rotators}.
The Hamiltonian of the rotator chain reads 
\begin{equation}\label{7.1}
H_\mathrm{CR} = \sum_{j=1}^N\big(\tfrac{1}{2}p_j^2 + V(\varphi_{j+1} - \varphi_j)\big)
\end{equation}
with periodic boundary conditions, $\varphi_{N+1} = \varphi_1$. At first glance we have only rebaptized $q_j$ as $\varphi_j$.
But the $\varphi_j$'s are angles and the $p_j$'s angular momenta. Hence the phase space is $(S^1\times\mathbb{R})^N$ with $S^1$ denoting the unit circle. The standard choice for $V$ is $V(\vartheta) =
-\cos \vartheta$, but in our context any $2 \pi$-periodic potential is admitted. The equations of motion are
\begin{equation}\label{7.2}
\frac{d}{dt} \varphi_j = p_j\,,\quad \frac{d}{dt}p_j = V'(\varphi_{j+1} - \varphi_j)- V'(\varphi_{j} - \varphi_{j-1})\,.
\end{equation}
Obviously angular momentum is locally conserved with the angular momentum current
\begin{equation}\label{7.3}
\mathcal{J}_1(j) = - V'(\varphi_{j} - \varphi_{j-1})\,.
\end{equation}
As local energy we define $e_j = \tfrac{1}{2}p_j^2 + V(\varphi_{j+1} - \varphi_j) $. Then $e_j$ is locally conserved, since
\begin{equation}\label{7.4}
\frac{d}{dt} e_j = p_{j+1} V'(\varphi_{j+1} - \varphi_{j})-p_jV'(\varphi_{j} - \varphi_{j-1})\,,
\end{equation}
from which one reads off the energy current
\begin{equation}\label{7.3a}
\mathcal{J}_2(j) = - p_j V'(\varphi_{j} - \varphi_{j-1})\,.
\end{equation}
For the angles, in analogy to the stretch,  one defines the phase difference $\tilde{r}_j = \Theta(\varphi_{j+1} -\varphi_j)$, where $\Theta$ is $2\pi$-periodic and $\Theta(x) = x$ for $|x| \leq \pi$. Because of the jump discontinuity the stretch is not conserved. A rotator chain has only two conserved fields.

To apply nonlinear fluctuating hydrodynamics one has to compute the Euler currents in local equilibrium.
Since there are two conserved fields the canonical equilibrium state reads
\begin{equation}\label{7.5}
\frac{1}{Z_N}\prod_{j= 1}^N \exp\big[ - \beta \big(\tfrac{1}{2}(p_j-u)^2 + V(\varphi_{j+1} - \varphi_j)\big)\big]  d\varphi_j dp_j\
\end{equation}
with $u$ the average angular momentum. Now
\begin{equation}\label{7.6}
\langle \mathcal{J}_1(j)\rangle_N = - \langle V'(\varphi_{j} - \varphi_{j-1})\rangle_N\,,\quad
 \langle \mathcal{J}_2(j)\rangle_N = - u\langle V'(\varphi_{j} - \varphi_{j-1})\rangle_N \,,
\end{equation}
average with respect to the canonical ensemble (\ref{7.5}). We claim that
\begin{equation}\label{7.7}
\lim_{N\to \infty}\langle V'(\varphi_{j} - \varphi_{j-1})\rangle_N = 0\,.
\end{equation}
For this purpose we expand in Fourier series as
\begin{equation}\label{7.8}
e^{-V(\vartheta)} = \sum_{m \in \mathbb{Z}}a(m) e^{-\mathrm{i} m\vartheta}\,,\quad f(\vartheta) = \sum_{m \in \mathbb{Z}} \hat{f}(m) e^{-\mathrm{i} m\vartheta}\,.
\end{equation}
Then, working out all Kronecker deltas from the integration over the $\varphi_j$'s, one arrives at
\begin{equation}\label{7.9}
\langle f(\varphi_{j+1} - \varphi_{j})\rangle_N  = \Big(  \sum_{m \in \mathbb{Z}}a(m)^N\Big)^{-1}
\sum_{m \in \mathbb{Z}}a(m)^N\Big(a(m)^{-1} \sum_{\ell \in \mathbb{Z}}\hat{f}(\ell -m)
 a(\ell)\Big)\,.
\end{equation}
Since $a(0) > |a(m)|$ for all $m \neq 0$,
\begin{equation}\label{7.10}
\lim_{N \to \infty} \langle f(\varphi_{j+1} - \varphi_{j})\rangle_N = a(0)^{-1} \sum_{\ell \in \mathbb{Z}}\hat{f}(\ell) a(\ell)   
= \frac{1}{Z_1} \int_{-\pi}^\pi d \vartheta f(\vartheta) e^{-\beta V(\vartheta)}\,.
\end{equation}
 For $f(\vartheta) = V'(\vartheta)$, the latter integral vanishes because of periodic boundary conditions in $\vartheta$.
 We conclude that both currents vanish on average.

As before we consider the infinite lattice with thermal expectation
 $\langle \cdot \rangle_{u,\beta}$  and form the equilibrium time correlations as
\begin{equation}\label{7.11}
S_{\alpha\alpha'}(j,t) = \langle g_\alpha(j,t)g_{\alpha'}(0,0) \rangle_{u,\beta} -
\langle g_\alpha(0) \rangle_{u,\beta}\langle g_\alpha'(0) \rangle_{u,\beta}\,,
\end{equation}
$\vec{g}(j,t) = \big(p_j(t),e_j(t)\big)$. $p_j$ is odd and $e_j$ is even under time reversal. Hence in the Green-Kubo formula
the cross term vanishes.  Thus, for large $j,t$, fluctuating hydrodynamics predicts
\begin{equation}\label{7.12}
S_{\alpha\alpha'}(j,t) = \delta_{\alpha\alpha'}(4\pi D_\alpha t)^{-1/2}f_{\mathrm{G}}((4\pi D_\alpha t)^{-1/2}j )\,,
\end{equation}
where $f_{\mathrm{G}}$ is the unit Gaussian. $D_\alpha$ is the diffusion coefficient of mode $\alpha$.
Of course, it can be written as a time-integral over the corresponding total current-current correlation,
but its precise value has to be determined numerically. This has been done for the standard choice 
$V(x) = -\cos x$, to which we specialize now. For $\beta = 1$, with a lattice size $N =500$, the diffusive peaks are well established at 
$t = 2000$ \cite{Ha14,DaDh14}. Energy diffusion has been confirmed much earlier \cite{GLPV00,GeSa00}. 

At low temperatures one finds a different, perhaps more interesting scenario. At zero temperature, there is the one-parameter family of ground states with $\varphi_j = \bar{\varphi}$, $p_j= 0$. When heating up, under the canonical equilibrium measure, the phase $\varphi_j$ jumps to $\varphi_{j+1}$ with a jump size $\mathcal{O}(1/\sqrt{\beta})$. 
Next we have to understand how the conservation of $\tilde{r}_j$ field is broken. In a pictorial language, the event that $\lvert\varphi_{j+1}(t) -\varphi_j(t)\rvert =\pi$ is called an umklapp for phase difference $\tilde{r_j}$ or an umklapp process to emphasize its dynamical character. At low temperatures a jump of size $\pi$ has a small probability of order $e^{-\beta\Delta V}$ with $\Delta V = 2$ the height of the potential barrier. Hence $\tilde{r}_j$ is locally conserved up to umklapp processes occurring at a very small frequency only. This can be measured more quantitatively by considering the average
\begin{equation}\label{7.13a}
\Gamma_\mathrm{uk}(t) = \sum_{j\in \mathbb{Z}}\big( \langle \tilde{r}_j(t)\tilde{r}_0(0)\rangle_{u,\beta} - \langle\tilde{r}_0\rangle_{u,\beta}^2\big)\,.
\end{equation}
At $\beta = 1$,  $\Gamma_\mathrm{uk}(t)$ decays exponentially due to umklapp. But at $\beta = 5$ the decay rate
is already very much suppressed, see \cite{DaDh14}.

In the low temperature regime it is tempting to use an approximation, where the potential $V(x) = -\cos x$ is Taylor expanded at the minimum $x=0$. But such procedure would underestimate the regime of low temperatures, as can be seen from the example of a potential, still with $\Delta V = 2$, but several shallow minima. The proper small parameter is $\beta^{-1}$ such that $\beta \Delta V > 1$. To arrive at an optimal low temperature hamiltonian, we first parametrize the angles $\varphi_1, \dots, \varphi_{N}$ through $r_j = \varphi_{j+1} -\varphi_j$ with $r_j \in [-\pi,\pi]$. To distinguish, we denote the angles in this particular parametrization by $\phi_j$. The dynamics governed by $H_{\mathrm{CR}}$ corresponds to periodic boundary conditions at $r_j = \pm \pi$. For a low temperature description we impose instead specular reflection, i.e., if $r_j = \pm\pi$, then $p_j$, $p_{j+1}$ are scattered to $p_j' = p_{j+1}$, $p_{j+1}' = p_j$. By fiat all umklapp processes are now suppressed, while between two umklapp events the CR dynamics and the low temperature dynamics are identical. The corresponding hamiltonian reads
\begin{equation}
H_{\mathrm{CR,lt}} = \sum_{j=1}^{N} \Big(\tfrac{1}{2} p_j^2 + \tilde{V}(\phi_{j+1} - \phi_j)\Big)
\end{equation}
with
\begin{equation}
\tilde{V}(x) = -\cos x \ \ \text{for} \ \ \lvert x \rvert \leq \pi\,,\qquad \tilde{V}(x) = \infty \ \ \text{for}\ \ \lvert x \rvert > \pi\,,
\end{equation}
periodic boundary conditions $\phi_{N+1} = \phi_1$ being understood. The pair $(\phi_j,p_j)$ are canonically conjugate variables. Note that as weights $\exp[-\beta H_{\mathrm{CR}}] = \exp[-\beta H_{\mathrm{CR,lt}}]$. Thus all equilibrium properties of the coupled rotators remain untouched.

The hamiltonian $H_\mathrm{CR,lt}$ is a variant of the hard collision model with square well potential as discussed before, see \cite{MeSp14}. The dynamics governed by $H_\mathrm{CR,lt}$ has three conserved fields, the stretch $r_j = \phi_{j+1} - \phi_j$, the momentum $p_j$, and the energy $e_j = \tfrac{1}{2}p_j^2 + \tilde{V}(r_j)$. Because of $\phi_1 = \phi_{N+1}$, one has $\sum_{j=1}^{N} r_j = 0$. The model is in the dynamical phase characterized by an even potential at zero pressure.

We claim that, for $\beta \Delta V > 1$, the CR equilibrium time correlations are well approximated by those of $H_\mathrm{CR,lt}$, provided the time of comparison is not too long. The latter correlations can be obtained within the framework of nonlinear fluctuating hydrodynamics. Thereby one arrives at fairly explicit dynamical predictions for the low temperature regime of the CR model.

One physically interesting information concerns the Landau-Placzek ratio at low temperatures. We use \eqref{51a}
and expand $R^{-1}$ in $1/\beta$. To lowest order it suffices to use the harmonic approximation, $  \tilde{V}(x)
=\tfrac{1}{2}ax^2$, $a = 1$ for the cosine potential. Using the formulas in Appendix A of \cite{Sp14} one arrives at the following value for the Landau-Placzek ratios, 
\begin{equation}\label{7.14}
r\mbox{-}r: \hspace{4pt} (2a\beta)^{-1}(1,0,1)\,,\quad p\mbox{-}p: \hspace{4pt}(2\beta)^{-1} (1,0,1)\,, \quad 
e\mbox{-}e: \hspace{4pt}
(2\beta)^{-1} (a\ell^2 , \beta^{-1}, a\ell^2   )\,.
\end{equation}
The correlations are small, order $\beta^{-1}$. For the stretch correlations there is no central peak, to this order, and for the energy correlations the central peak is down by a factor $\beta^{-1}$ relative to the sound peaks. 

To have a unified picture we add $ \tilde{r}_j$ to the list of  fields of physical interest. At high temperatures
$ \tilde{r}_j$ is not conserved and one has diffusive spreading of the conserved fields.
At low temperatures
$ \tilde{r}_j$ is conserved up to small errors and the conventional three-peak structure, including universal shape functions, results. For extremely long times umklapp processes will happen and one expects that they force  a cross over to the Gaussian scaling (\ref{7.12}). The precise dynamical structure of such cross over still needs to be investigated.\medskip\\
\textbf{Nonlinear Schr\"{o}dinger equation on a lattice}.
A further example with a dynamically distinct low temperature phase is  the nonlinear Sch\"{o}dinger equation on the one-dimensional lattice. In this case the lattice field is $\psi_j \in \mathbb{C}$, for which real and imaginary part are the canonically conjugate fields. The Hamiltonian reads
\begin{equation}\label{7.15}
H = \sum_{j=1}^N \big(\tfrac{1}{2}|\psi_{j+1} - \psi_j|^2 + \tfrac{1}{2}g |\psi_j|^4 \big)
\end{equation}
with periodic boundary conditions and coupling $g > 0$.  The sign of the hopping term plays no role, since it can be switched through  the gauge transformation $\psi_j \leadsto
\mathrm{e}^{\mathrm{i}\pi j}\psi_j$. The chain is non-integrable and the locally conserved fields are the number density
$\rho_j = |\psi_j|^2$ and the local energy $e_j = \tfrac{1}{2}|\psi_{j+1} - \psi_j|^2+  \tfrac{1}{2}g |\psi_j|^4$.  
Hence the canonical equilibrium state is given by
\begin{equation}\label{7.16}
Z^{-1}\mathrm{e}^{-\beta(H - \mu \mathsf{N})}\prod_{j=1}^N d\psi_jd\psi_j^*\,,\quad \mathsf{N} =   \sum_{j=1}^N |\psi_j|^2\,,
\end{equation}
with  the chemical potential $\mu$. We assume $\beta >0$. But  also negative temperature states, in the microcanonical ensemble, have been studied \cite{ILLP13,IPP13}. Then the dynamics is dominated by a coarsening process mediated through breathers.
In equilibrium, the $\psi$-field has high spikes at random locations embedded in a low noise background, which is very different from the positive temperature states considered here. For them
the density and energy currents are symbolically of the form $\mathrm{i}(z - z^*)$, hence
 their thermal average vanishes. Both fields are expected to have diffusive transport. In fact, this is
 confirmed by MD simulations \cite{MeSp15b}. They also show Gaussian cross-correlations, which is possible since density and energy are both even under time reversal. In the previous studies \cite{Iu12}
 transport coefficients have been measured in the steady state set-up.

To understand the low temperature phase, it is convenient to transform to the new canonical pairs $\rho_j, \varphi_j$ through
\begin{equation}\label{7.17}
\psi_{j} = \sqrt{\rho_j}\,\mathrm{e}^{\mathrm{i}\varphi_j}\,.
\end{equation}
In these variables the Hamiltonian becomes
\begin{equation}
\label{eq:polarHamiltonian}
H = \sum_{j=1}^N  \big(- \sqrt{\rho_{j+1}\,\rho_j} \cos(\varphi_{j+1} - \varphi_j) + \rho_j + \tfrac{1}{2}g\rho_j^2 \big)\,.
\end{equation}
The equations of motion read then
\begin{equation}\label{7.17a}
\partial_t \varphi_j = -\partial_{\rho_j} H\,, \quad \partial_t \rho_j = \partial_{\varphi_j} H\,.
\end{equation}
$\varphi_j$ takes values on the circle $S^1$ and $\rho_j \geq 0$. From the continuity of $\psi_j(t)$ when moving through the origin, one concludes that at $\rho_j(t)= 0$ the phase jumps from $\varphi_j(t) $ to $ \varphi_j(t) + \pi$.

One recognizes the similarity to the coupled rotators \eqref{7.1}. But now the equilibrium measure carries a nearest neighbor coupling. 
For $\mu > 0$, in the limit $\beta \to \infty$ the canonical measure converges to the one-parameter family of ground states with $\rho_j = \bar{\varphi} =  \mu/g$, $\varphi_j
= \bar{\varphi}$ with $\bar{\varphi}$  uniformly distributed on $S^1$.   At low temperatures the field of phase differences $\tilde{r}_j$ is approximately conserved. 
The low temperature hamiltonian is constructed in such a way that the equilibrium ensemble remains unchanged while all umklapp processes are suppressed. To achieve our goal we follow verbatim the CR blueprint. The phases are parametrized such that $\varphi_{j+1} - \varphi_j$ lies in the interval $[-\pi,\pi]$ and this particular parametrization denoted by $\phi_j$. Umklapp is a point at the boundary of this interval. Now $(\phi_j,\rho_j)$ are a pair of canonically conjugate variables, only $\rho_j \geq 0$ instead of $p_j \in \mathbb{R}$. Thus the proper low temperature hamiltonian reads
\begin{equation}
H_{\mathrm{lt}} = \sum_{j=0}^{N-1} \Big( \sqrt{\rho_{j+1}\,\rho_j} \, U(\phi_{j+1} - \phi_j) + V(\rho_j) \Big)\,,
\end{equation}
where
\begin{equation}
U(x) = - \cos(x) \ \ \text{for}\ \ \lvert x \rvert \leq \pi \,,\qquad U(x) = \infty \ \ \text{for}\ \ \lvert x \rvert > \pi\,,
\end{equation}
and 
\begin{equation}
V(x) =  x + \tfrac{1}{2} g\, x^2 \ \ \text{for}\ \ x \geq 0\,,\qquad V(x) = \infty \ \ \text{for}\ \ x < 0\,.
\end{equation}

The low temperature Hamiltonian has a nearest neighbor coupling, which complicates the scheme through 
which the $\vec{G}$ matrices are determined \cite{MeSp15b}. Progress is achieved through the miraculous identity
$\vec{\mathsf{j}} = (\mu,P,\mu P)$ for the Euler currents. The $\vec{G}$ coefficients are evaluated at $P = 0$. As for a generic anharmonic chain, one finds that
$G^0_{00} = 0$ and $G_{11}^{1} \neq 0$. Thus the heat peak is predicted to be L\'{e}vy 5/3
and the sound peaks to be KPZ. The sound peak for the density-density correlation was first observed in 
 \cite{KuLa13} using $k,\omega$ space, see also \cite{KuHuSp14}. In \cite{MeSp15b} we use normal mode representation, as explained in this article. The sound peaks fit nicely with KPZ, but the normalized heat peak is very broad and noisy,
 still with a shape not unlikely  L\'{e}vy 5/3.

\end{document}